\documentclass{article}

    \PassOptionsToPackage{numbers, compress}{natbib}
\usepackage{natbib} 
    \usepackage[final]{neurips_2025}

\usepackage[utf8]{inputenc} 
\usepackage[T1]{fontenc}    
\usepackage{hyperref}       
\usepackage{url}            
\usepackage{booktabs}       
\usepackage{amsfonts}       
\usepackage{nicefrac}       
\usepackage{microtype}      
\usepackage{xcolor}         

\usepackage{microtype}
\usepackage{graphicx}
\usepackage{subfigure}
\usepackage{colortbl}
\usepackage{enumitem}  
\usepackage{amsmath}
\usepackage{amssymb}
\usepackage{mathtools}
\usepackage{amsthm}
\usepackage{multirow,hhline}
\usepackage{caption}
\usepackage{algorithm}
\usepackage{algpseudocode}
\usepackage{amsmath}
\usepackage{bm}
\usepackage{makecell}
\usepackage{wrapfig}
\usepackage{amsthm}
\usepackage{booktabs}
\usepackage{xcolor}

\usepackage{tcolorbox}
\usepackage{listings}
\usepackage{xcolor}

\newtheorem{proposition}{Proposition}[section]

\definecolor{GoogleGreen}{RGB}{60,186,84}
\definecolor{GoogleRed}{RGB}{219,50,54}
\newcommand{\red}[1]{\textcolor{GoogleRed}{#1}} 
\newcommand{\green}[1]{\textcolor{GoogleGreen}{#1}}

\usepackage[capitalize,noabbrev]{cleveref}

\usepackage[textsize=tiny]{todonotes}

\newcommand{\model}{{SEEK}}

\title {Enhancing LLM Watermark Resilience Against Both Scrubbing and Spoofing Attacks}

\author{%
  Huanming Shen\textsuperscript{1,2}\thanks{Equal contribution}  \\
  \texttt{huanmingshen@std.uestc.edu.cn} \\
  \And
  Baizhou Huang\textsuperscript{1}\footnotemark[1] \\
  \texttt{hbz19@pku.cn} \\
  \And
  Xiaojun Wan\textsuperscript{1}\thanks{Corresponding author} \\
  \texttt{wanxiaojun@pku.cn} \\
  \AND
  \textsuperscript{1}\textnormal{Wangxuan Institute of Computer Technology, Peking University} \\
  \textsuperscript{2}\textnormal{University of Electronic Science and Technology of China}
}

\definecolor{mygray}{gray}{.91}

\begin{document}
\maketitle


\begin{abstract}
Watermarking is a promising defense against the misuse of large language models (LLMs), yet it remains vulnerable to scrubbing and spoofing attacks.
This vulnerability stems from an inherent trade-off governed by watermark window size: 
smaller windows resist scrubbing better but are easier to reverse-engineer, enabling low-cost statistics-based spoofing attacks. 
This work expands the trade-off boundary by introducing a novel mechanism, equivalent texture keys, where multiple tokens within a watermark window can independently support the detection. 
Based on the redundancy, we propose a watermark scheme with \textbf{S}ub-vocabulary decomposed \textbf{E}quivalent t\textbf{E}xture \textbf{K}ey (\textbf{SEEK}). 
It achieves a Pareto improvement, increasing the resilience against scrubbing attacks without compromising robustness to spoofing. 
\end{abstract}


\section{Introduction}

Recent advances in large language models~\citep{lm-4,lm-5,lm-6,llama,llama2,chatgpt,gpt4} have significantly improved the realism of AI-generated text, making it increasingly difficult to distinguish from human-written content. However, this progress also raises concerns about potential misuse, including disinformation dissemination~\citep{woodcock2023wikipedia}, automated phishing~\citep{stokel2022ai}, and academic integrity~\citep{pudasaini2024survey}. In response, detecting LLM-generated text has become an active area of research, with growing interest in techniques for enhancing provenance tracking and content verification.


One proactive approach involves watermarking, wherein the LLM's output distribution is subtly altered to embed an imperceptible signature within the generated text~\citep{KWG1,kuditipudi2023robust,fairoze2023publicly,zhao2023provable}. This signature can subsequently be verified by a detector equipped with a secret key held by the LLM provider. Compared to post-hoc detection methods~\citep{gehrmann2019gltr,mitchell2023detectgpt,hovy2016enemy, zellers2019defending}, watermarking offers notable advantages, including the preservation of output quality and a provably low false positive rate (below $10^{-3}$)~\citep{hans2401spotting}. However, watermarking remains vulnerable to two types of adversarial attacks: 1)~\textbf{scrubbing attacks} which generate a semantically equivalent paraphrase to disturb watermark patterns, thereby bypassing the watermark detection and  2)~\textbf{spoofing attacks} which mimic the watermark patterns to inject target watermarks into harmful text, fabricating content that appears to be generated by the victim LLM.

A widely acknowledged perspective in existing research \citep{steal,KWG2,liu2023unforgeable,pang2024attacking,Liu2023ASI} is that distribution-modifying watermarking exhibits an inherent trade-off between scrubbing robustness and spoofing robustness. Among such methods, KGW~\citep{KWG1} stands out as a representative approach, introducing a family of $h$-gram statistical watermarks~\citep{hou2023semstamp,zhao2023provable,KWG2}. They modify the language model’s next token distribution at each step based on the preceding $h$ tokens, referred to as the watermark window. 
\begin{wrapfigure}{r}{0.45\textwidth}  %
    \centering
    \includegraphics[width=0.45\textwidth]{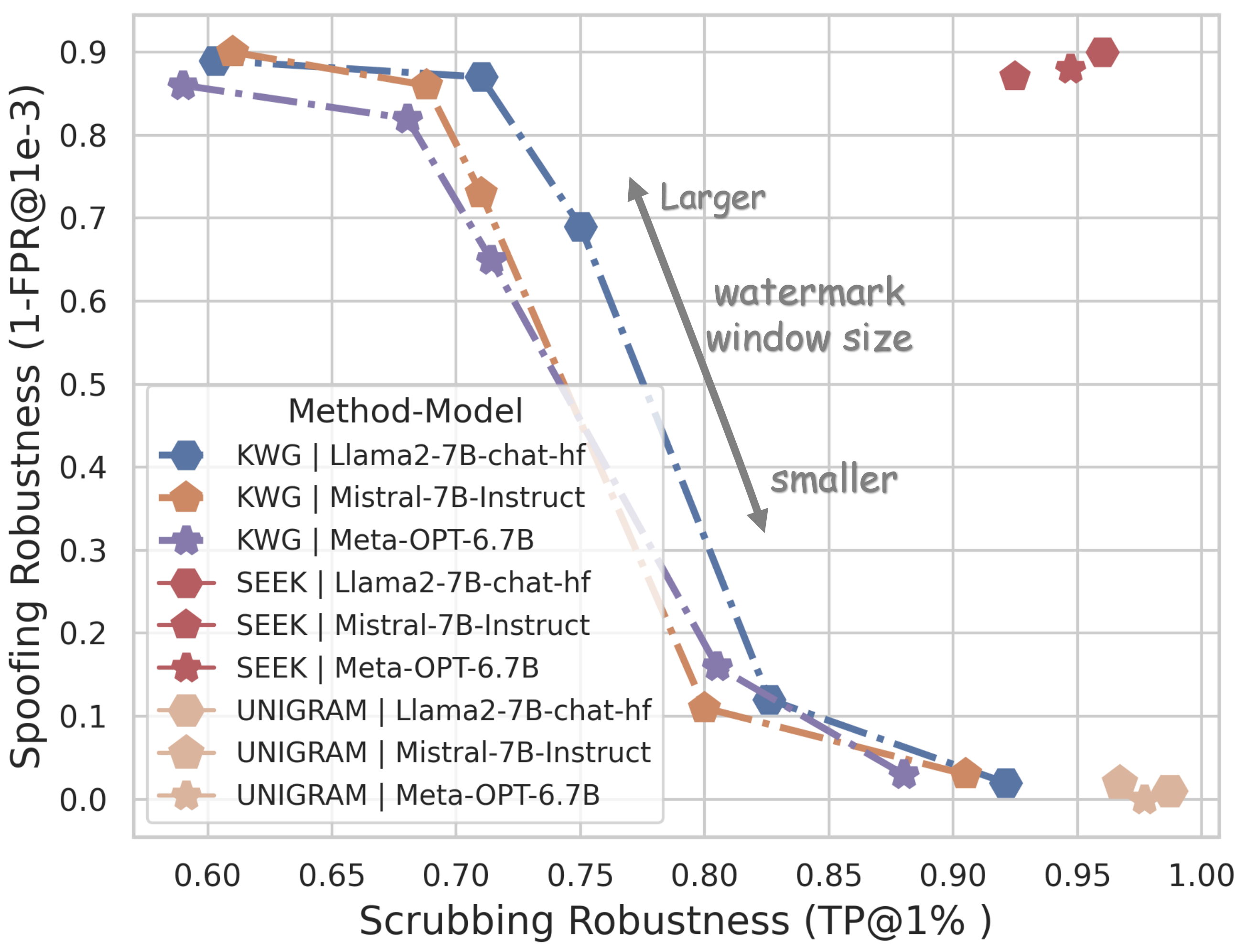}
    \vspace{-1.5em}
    \caption{\small Performance of different schemes under scrubbing and spoofing attacks. Varying the watermark window size induces a trade-off between scrubbing and spoofing robustness. Scrubbing robustness is evaluated using DIPPER~\citep{dipper} on the C4-RealNewsLike dataset. Spoofing robustness is evaluated using statistics-based attacks~\citep{steal} on the Dolly-CW dataset. Ours achieves improved robustness on both axes, reaching Pareto optimality.}
    \label{fig:toutu}
    \vspace{-1.0em}
\end{wrapfigure}
A larger watermark window requires longer unaltered text segments for reliable detection, thus increasing susceptibility to localized edits, i.e., scrubbing. Therefore, to maintain resilience against scrubbing attacks like paraphrasing, these methods typically restrict the watermark window size to fewer than four tokens. However, the small window setting also introduces a structural limitation that can be exploited by spoofing attacks. Recent study~\citep{steal} has revealed that by collecting moderate samples from the watermarked LLM, adversaries can statistically learn the mapping from watermark windows to corresponding watermarked outputs. This enables the generation of high-quality spoofed text that closely mimics the characteristics of genuinely watermarked text. 
In essence, the efficacy of such adversarial attacks is not primarily due to their sophistication, but rather the inherent limitations of conventional schemes that force a direct compromise between scrubbing and spoofing robustness via the singular control of the watermark window size, as shown in Figure~\ref{fig:toutu}. This presents a compelling research question: 


\textit{How can watermarking schemes be enhanced to simultaneously ensure robustness against scrubbing and resistance to spoofing attacks?}

To address this question, we investigate an overlooked factor influencing robustness against scrubbing attacks beyond the watermark window size. We identify a phenomenon of equivalent texture keys, where multiple tokens within a watermark window can independently support the detection of the watermark pattern at the subsequent timestep. This redundancy enables a path towards Pareto improvements: by increasing the number of equivalent texture keys, scrubbing robustness can be significantly enhanced even when employing large watermark windows, without compromising resistance to spoofing attacks. 

Building on this insight, our initial approach increases the prevalence of equivalent texture keys by reducing the output space of the hash function typically used in watermarking. However, recognizing that a smaller hash output space could potentially degrade text quality, we introduce our final watermark scheme, \textbf{SEEK} (\textbf{S}ub-vocabulary decomposed \textbf{E}quivalent t\textbf{E}xture \textbf{K}ey). SEEK ingeniously mitigates this quality concern by decoupling the watermark construction process across disjoint sub-vocabularies. This allows for a high density of equivalent texture keys, bolstering robustness, while simultaneously preserving the quality of the generated text.
Our contributions are as follows:

\setlist[itemize]{leftmargin=*}
\begin{itemize}
    \item We reveal that equivalent texture keys, where multiple tokens within a watermark window independently support detection, are critical for enhancing scrubbing robustness under large windows. 
    
    \item We propose a novel scheme that improves robustness by leveraging the equivalent texture key mechanism, while simultaneously guaranteeing text quality through the decomposition of watermark construction across disjoint sub-vocabularies. 
    
    \item The experiments demonstrate that SEEK outperforms KGW-family baselines, achieving substantial improvements in scrubbing and spoofing robustness across datasets, establishing a new Pareto frontier, yielding spoofing robustness gains of +88.2\%/+92.3\%/+82.0\% on Dolly-CW/MMW-BookReports/MMW-FakeNews and scrubbing robustness gains of +10.2\%/+13.4\%/+8.6\% on WikiText/C4/LFQA compared to KGW-Min with 4-gram watermark window. 
\end{itemize}

\section{Related work}

\textbf{LLM Watermarking technology.} Watermarking natural language faces challenges due to the discrete and symbolic nature of text~\citep{stefan2000information}. Early methods employed rule-based techniques such as syntactic restructuring~\citep{atallah2001natural}, synonym substitution~\citep{topkara2006hiding}, and paraphrasing~\citep{atallah2002natural}. Foundational pioneering work in LLM watermarking includes KGW~\citep{KWG1,KWG2} and AAR~\citep{arr} builds watermarks on statistics for a low and provable false positive rate. Specifically, watermark detection was framed as a hypothesis testing problem, where the alternative hypothesis models text generation from a modified distribution. The KGW method partitions the vocabulary into a "green" list and a "red" list to modify the next token's distribution,  which significantly influenced numerous follow-up watermark works~\citep{kuditipudi2023robust,mao2024watermark, sweet,zhao2023provable,liu2023unforgeable,lu2024entropy}. Our work focuses on the security properties of KGW-family watermarking schemes.


\textbf{Scrubbing attacks.} Scrubbing attacks fundamentally stem from text editing operations. Early studies~\cite{zhao2023provable,kuditipudi2023robust, KWG1} evaluated the robustness of watermarking schemes using simple perturbations such as random word deletion or substitution. In addition, some work~\citep{KWG2,zhao2023provable} has explored manual paraphrasing by human annotators to assess watermark removal under high-quality rewriting. Another line of work aims to develop black-box scrubbing attacks, like ~\citet{zhang2023watermarks} and \citet{sadasivan2023can} propose recursive paraphrasing pipelines, while \citet{huang2024b} formulate the scrubbing task as a constrained optimization problem. \citet{dipper} propose a systematic approach to evade AI-generated text detection by training a paraphrasing model.

\textbf{Spoofing attacks.} Early work~\citep{sadasivan2023can,hou2023semstamp,ghosal2023towards,gu2023learnability} has demonstrated proof-of-concept spoofing against simple $h=1$ watermark schemes known as KGW-Left~\citep{KWG1}. \citet{gu2023learnability} propose a distillation-based spoofing method, which requires considerable training overhead. \citet{zhang2024large} formalize spoofing as a mixed-integer programming problem; however, their method depends on access to the model's full vocabulary and tokenizer. In contrast, \citet{steal} proposes a highly effective statistics-based spoofing attack that approximates the watermark distribution via token frequency analysis, without any model-specific access. By learning watermark prefix \( p_w(x_n|x_{n-h:n-1}) \) and non-watermark prefix \( p_m(x_n|x_{n-h:n-1}) \), they successfully spoof KGW-Min~\cite{KWG2} for under \$50.

\section{Preliminaries}

\subsection{Watermarking in LLMs}
\textbf{Large Language models.} Following the notation from prior work~\citep{wu2023resilient,hou2024k,huang2024b,mao2024watermark,kuditipudi2023robust}, we define the generation process of a LLM. Let \(\mathcal{M}\) denote a pretrained LLM with a vocabulary \(\mathcal{V}\), typically comprising over 50{,}000 tokens (i.e., \( |\mathcal{V}| > 50{,}000 \))~\citep{kuditipudi2023robust}.  Given an input \( x_{1:n-1} = \{x_1, x_2, \dots, x_{n-1}\} \), the model generates a sequence by computing token-wise logit scores \(\ell_n\) and sampling tokens from the output distribution \( x_n \sim P_\mathcal{M}(\cdot \mid x_{1:n-1}) \) in an autoregressive fashion.

\textbf{LLM watermarking.} Watermarking alters the origin model‘s generation distribution $P_M(\cdot | x_{1:n-1})$, yielding a modified distribution \( P_W(\cdot | x_{1:n-1}) = \mathcal{T}(P_M(\cdot | x_{1:n-1}),\theta_n)\), where $\theta_n$ denotes the cipher for the current step. KGW-family methods typically conduct the modification $\mathcal{T}$ through a random partition of the vocabulary seeded by the cipher $\theta_n$. Specifically, a subset \( G \subseteq \mathcal{V} \) of predefined size \( \gamma |\mathcal{V}| \) is sampled as the green list, while the rest is termed as the red list $R=\mathcal{V} \setminus G$. Then a positive bias \( \delta \) is added to the logits of tokens in the green list, shifting the distribution to favor sampling green list tokens. Specifically, the modified logit of token $t$, denoted by $\hat{\ell}_n[t]$, is formulated as:
\begin{equation}\label{eq:equation1}
\hat{\ell}_n[t]=\ell_n[t]+\delta\cdot\mathbf{1}_{G}(t)
\end{equation}
where $\mathbf{1}(\cdot)$ is the indicator function. 
During the detection phase, A one-sided significance test is conducted to distinguish between watermarked and non-watermarked text~\citep{KWG1,huang2024b,kuditipudi2023robust}. Formally, given a candidate sequence $\tilde{x}_{1:N} = \{x_1, \dots, x_N\}$, the detection problem is framed as a z-statistic test with tokens in all timesteps as i.i.d. samples. We reconstruct the green list at each step $n$ using the cipher $\theta_n$, and count the green tokens in the candidate sequence as the test statistic, from which the $z$-score is computed. If $z$-score exceeds a certain threshold, we conclude that the watermark is embedded in $x$ and thus generated by the LLM $\mathcal{M}$. Notably, this test fully depends on the cipher $\theta_n$ in the candidate, rendering watermark detection infeasible if it is disturbed. \citet{KWG2} proposed a technique called self-seeding to incorporate the next token  \( x_{n} \) as part of the watermark window, which extends the watermark window by one token. The self-seeding mechanism and the watermark detection process are described in detail in Appendix~\ref{sec:self-seeding} and Appendix~\ref{sec:detection}, respectively.


\textbf{Derivation of the cipher $\theta_n$.}
The cipher $\theta_n=\xi\cdot\zeta$ consists of two components: a \textit{secret key} $\xi$ held by the service provider and a \textit{texture key}~\citep{wu2023resilient} $\zeta$ derived from the watermark window. Specifically, given a hash function $H$ with a hash space of $\{1,2,3,...,d\}$, a hash signature $I=\{H(x_{n-k})|1\leq k\leq h\}$ of current watermark window is firstly calculated. Then an aggregation function $f: \mathbb{N}^{h} \to \mathbb{N}$ is applied on the signature to generate the texture key $\zeta$. Among several variants of the aggregation function, the $f=$\textit{Min} variant exhibits superior robustness~\citep{ghosal2023towards,liu2024survey}, which is formulated as:
\begin{equation}\label{eq:seed}
\theta_n =\zeta\cdot\xi=f(I)\cdot \xi=\operatorname{Min} I \cdot\xi =  \operatorname{Min}\left\{ H(x_{n-h})\cdot\xi, \dots, H(x_{{n-1}})\cdot\xi)  \right\}
\end{equation}
We also present other aggregation function $f$ variants in detail in Appendix~\ref{sec:KGW_Variants}. 

\subsection{Threat Models}
Following previous works \citep{steal,kuditipudi2023robust,pang2024strength,huang2024b}, the threat models involve two primary entities: \textit{victim} and \textit{attacker}. The victim operates a special language model $\mathcal{M}$ protected by watermarking and provides services externally via an API. The attacker aims to compromise the watermark detection algorithm through either scrubbing attacks or spoofing attacks. Under standard security assumptions (i.e., Kerckhoffs's principle \citep{kerckhoffs1883cryptographie}), we follow \cite{wu2024bypassing, steal} to assume the attacker has complete knowledge of the watermarking scheme, and can query the victim model via an API interface.

\textbf{Scrubbing’s objective.}
Scrubbing attacks refer to adversarial attempts to remove watermarks from protected text. Formally, given a watermarked response, an attacker aims at producing a semantically equivalent paraphrased output classified as non-watermarked text by the victim's detector. 

\textbf{Spoofing’s objective.} The objective of a spoofing attack is to construct a watermarking model \( \mathcal{M}' \) that mimics the text distribution of the victim model \( \mathcal{M} \). Subsequently, the attacker is able to generate a malicious message that is classified as watermarked text by the victim's detector.


\section{Method}
\subsection{Motivation: Pareto Improvements via Hash Collisions}
\label{sec:init-method}
A robust watermarking algorithm must not only accurately distinguish watermarked text but also maintain detection reliability under various adversarial perturbations. However, a fundamental trade-off exists between robustness against scrubbing attacks and resistance to spoofing attacks as shown in Figure \ref{fig:toutu}, which poses a significant challenge to the practical deployment of watermarking methods.

\begin{figure}[!ht]
    \centering
    \begin{minipage}{0.51\textwidth}
        \centering
\includegraphics[width=0.95\linewidth]{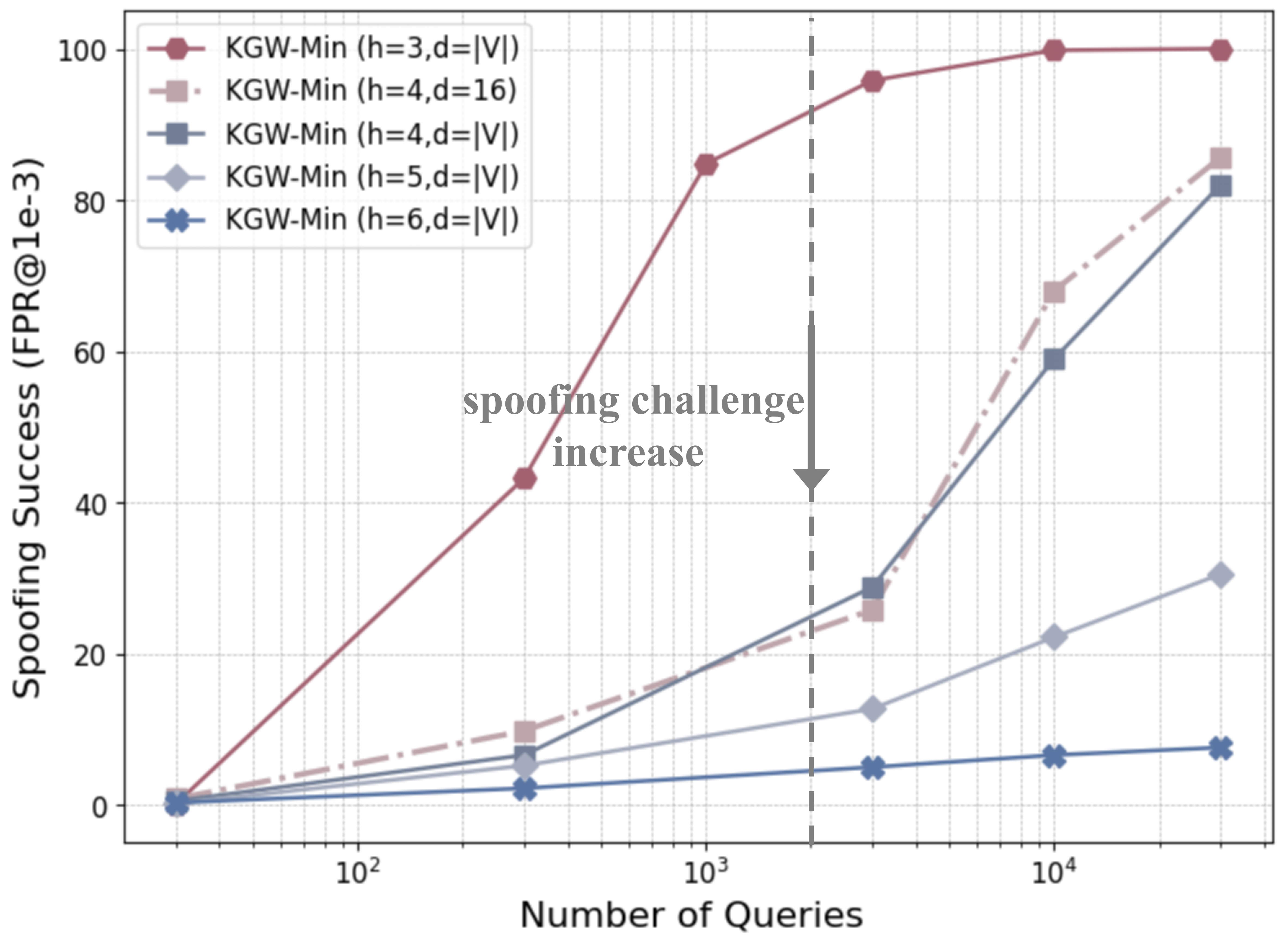}
        \vspace{-0.5em}
        \caption{\small Performance of different watermark window schemes under spoofing attack. The attack is conducted using 500 malicious texts generated by Dolly-CW, targeting a calibrated detector under FPR of 0.1\%. }
        \label{fig:sparsity}
    \end{minipage}\hfill
    \begin{minipage}{0.46\textwidth}
        \centering
        \captionof{table}{\small Comparison of the performance of various watermarking schemes against the ChatGPT scrubbing attack on the C4-Realnewslike dataset. $h$ represents the watermark window size, and $d$ indicates the cardinality of the hash space.
        }
        \resizebox{0.9\linewidth}{!}{
        \begin{tabular}{c|c|c}
        \bottomrule[1.5pt]
        \rowcolor{mygray}\textbf{Schemes}  &   \textbf{TP@1\%} $\uparrow$ & \textbf{TP@5\%} $\uparrow$ \\
        \toprule[0.75pt]
        \multicolumn{3}{c}{\textit{Origin}}\\
        \midrule
        KGW-Min (h=$4$, d=$|\mathcal{V}|$) & 73.5 & 86.5\\
        KGW-Min (h=$8$, d=$|\mathcal{V}|$) & 49.5 & 70.7 \\
        KGW-Min (h=$16$, d=$|\mathcal{V}|$) & 43.1 & 58.8 \\ 
        KGW-Min (h=$32$, d=$|\mathcal{V}|$) &32.4  &47.6  \\ 
        \midrule
        \multicolumn{3}{c}{\textit{With Equivalent Texture Keys}}\\
        \midrule
        UNIGRAM (h=$0 $, d=$1$) &91.8 & 94.9\\
        KGW-Min (h=$4 $, d=$16$) &80.3& 90.2\\
        KGW-Min (h=$8 $, d=$16$) &78.9& 87.8\\
        KGW-Min (h=$16 $, d=$16$) &72.9 & 85.9\\
        KGW-Min (h=$32 $, d=$16$)  &53.4 & 74.6\\
        \bottomrule[1.5pt]
        \end{tabular}
        }
        \label{tab:multi-texture-key}
    \end{minipage}
    \vspace{-1em}
\end{figure}

\label{sec:motivation}
\textit{\textbf{Finding I}: Statistical spoofing attacks suffer from sparsity: as the watermark window increases, the sample complexity grows exponentially, thereby reducing spoofing success rates.}

The efficacy of statistics-based spoofing~\citep{steal} stems from the statistical analysis of the watermark window, enabling the attacker to estimate the mapping between watermark windows and corresponding watermarked tokens. Intuitively, the complexity of such attacks is determined by the diversity of watermark windows. As shown in Figure~\ref{fig:toutu} and~\ref{fig:sparsity}, increasing the size of the watermark window leads to a substantial decline in spoofing success rates. This observation aligns with the sparsity challenge highlighted in~\citep{steal}, where sample complexity in statistical spoofing attacks scales as \(\mathcal{O}(|\mathcal{V}|^h)\). 

\textit{\textbf{Finding II}: For watermarking schemes with the same window size, reducing the cardinality of the hash space improves robustness against scrubbing attacks without compromising spoofing resistance.}

We investigate a variant of KGW that operates over a reduced hash space, thereby increasing the likelihood of hash collisions\footnote{A hash collision occurs when two tokens in a watermark window yield same hash value under $H(\cdot)$.} when deriving texture keys in Eq \ref{eq:seed}. 
As shown in Table~\ref{tab:multi-texture-key}, this variant achieves markedly gains in robustness against scrubbing attacks while maintaining watermark window size. We attribute the improvement to the increased frequency of hash collisions in the watermark window, which enables multiple tokens to independently reconstruct texture key $\zeta$ for valid detection. 

Consider KGW-MIN as an example, where texture key $\zeta=\operatorname{Min}\left\{ H(x_{n-h}), \dots, H(x_{{n-1}}))\right\}$. In the presence of frequent hash collisions, multiple tokens $x_j$ in the watermark window may satisfy $H(x_j)=\operatorname{Min} I$, enabling each to independently support watermark detection. We refer to this phenomenon as the emergence of \textit{Equivalent Texture Keys}. This redundancy enhances robustness: an attacker now needs to disturb all equivalent texture key tokens to remove the watermark. The challenge of scrubbing attacks towards our proposed variant can be formulated as follows:
\begin{proposition}
Given a hash function with space dimension $d$ and a watermark window size $h$, the probability of a hash collision occurring can be approximated by $p(h, d) \geq 1 - e^{\frac{-h(h-1)}{2d}}$.
\end{proposition}
\begin{proposition}
The expected number of tokens that must be erased to eliminate the watermark is expressed as $\sum_{m=1}^d \frac{h}{d} \left( \frac{d - m + 1}{d} \right)^{h-1}$, which is monotonically non-increasing in $d$ (strictly decreasing for $h\geq2$).
\end{proposition}

Formal proofs and corresponding visualizations for these propositions are presented in Appendix~\ref{sec:method}. From this perspective, UNIGRAM~\citep{ren2023robust}, a scheme known for its exceptional resilience to scrubbing, can also be interpreted as a special case of our proposed variant with the minimal hash space of cardinality $d=1$. On the other hand, reducing the hash space size does not influence the watermark window size, thereby maintaining similar spoofing resistance\footnote{However, if $d$ is reduced to 1, KGW-Min degenerates into a UNIGRAM scheme.}. Experimental results in Figure~\ref{fig:sparsity} indicate that the success rate of spoofing attacks is not affected by $d$. 

Taken together, these two findings underscore a key insight: the introduction of equivalent texture keys yields a form of Pareto improvement, enhancing robustness against scrubbing without sacrificing spoofing resistance, offering a promising pathway toward more practical watermark designs.

\subsection{\model: Sub-vocabulary Decomposed Equivalent Texture Keys}
\label{sec:seek}
\begin{wrapfigure}{r}{0.45\textwidth}  %
    \vspace{-1.5em}  
    \centering
    \includegraphics[width=0.45\textwidth]{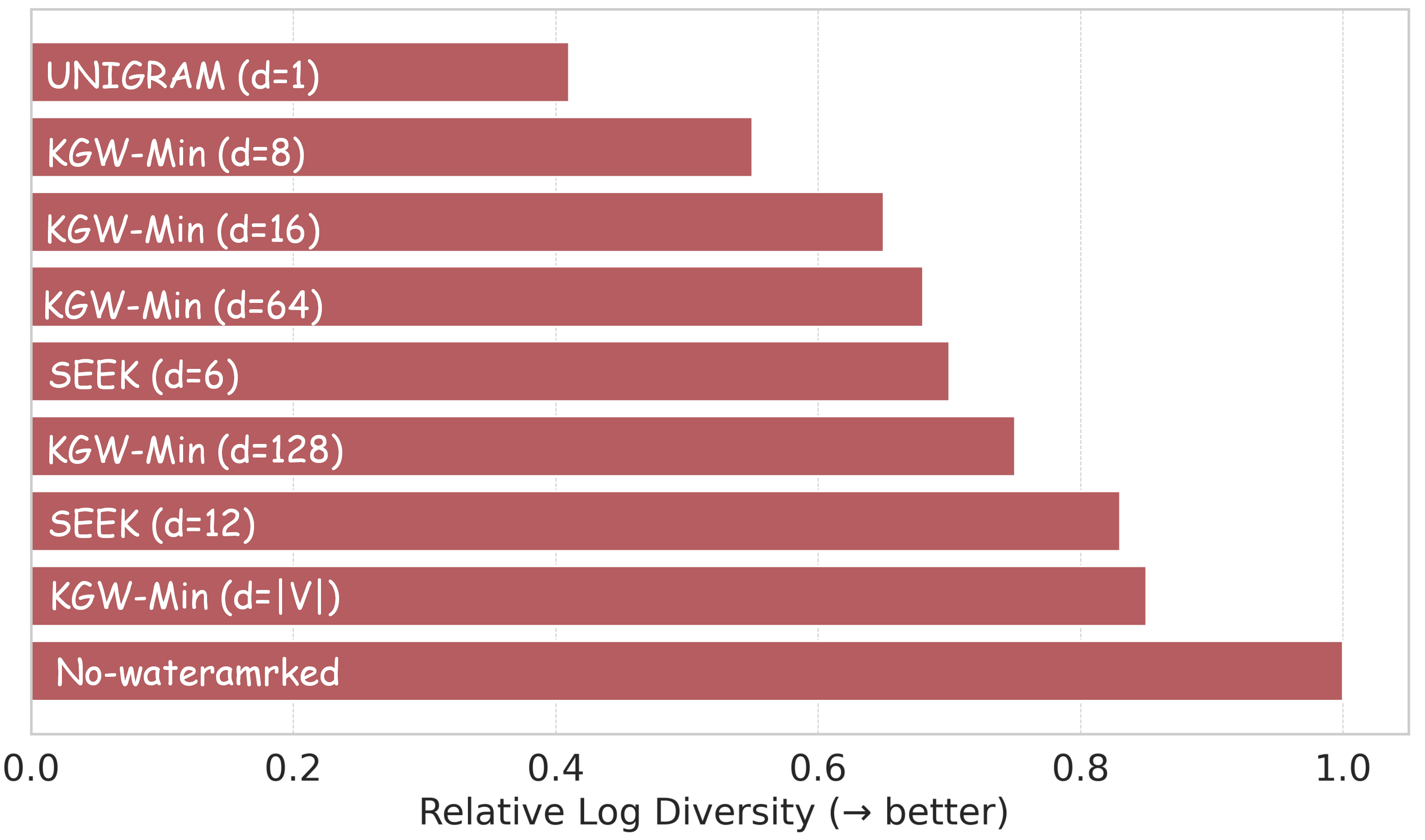}
    \vspace{-1.2em}
    \caption{\small An analysis of generation quality across different schemes on a subset of C4 by Log Diversity metric, relative to the unwatermarked text.}
    \label{fig:motivation3}
    \vspace{-2em}
\end{wrapfigure}

While reducing the size of the hash space yields significant improvements in watermark robustness, it inadvertently restricts the diversity of green-red list partitions. In the KGW-Min scheme, the diversity is upper bounded by \(\mathcal{O}(d)\), which severely degrades the text quality. As illustrated in Figure~\ref{fig:motivation3}, we observe a clear degradation in generation quality with the decrease of cardinality $d$. Prior research~\citep{pang2024no, Liu2023ASI} has also raised concerns that reducing the number of possible partitions may increase the risk of underlying pattern learning. In this section, we seek to harness the potential of equivalent texture keys while mitigating their adverse impact on generation quality.

We first revisit the variant proposed in the previous section.
Given the hash function $H(\cdot)$ maps to $\{1,2,...,d\}$, the cipher derivation in Equation~\ref{eq:seed} can be reformulated as follows:
\begin{equation}\label{eq:reformulation}
\forall i\in\{1,2,...,d\},\,\theta_n^{i}=i\cdot\xi\quad\quad
\theta_n =\operatorname{Min} \{\theta_n^{i}|i\in I\}
\end{equation} 
Since each $\theta_n^{i}$ corresponds to one unique green list partition of the vocabulary, $\operatorname{Min} \{\theta_n^{i}|i\in I\}$ essentially conducts a selection over all possible partitions, which is thereby upper bounded by the hash space size $d$. This formulation reveals that the reason for the inferior diversity of green-red list partitions stems from the dependency on one single cipher. 

\begin{figure}[t]
    \vspace{-1em}  
    \centering
    \includegraphics[width=\textwidth]{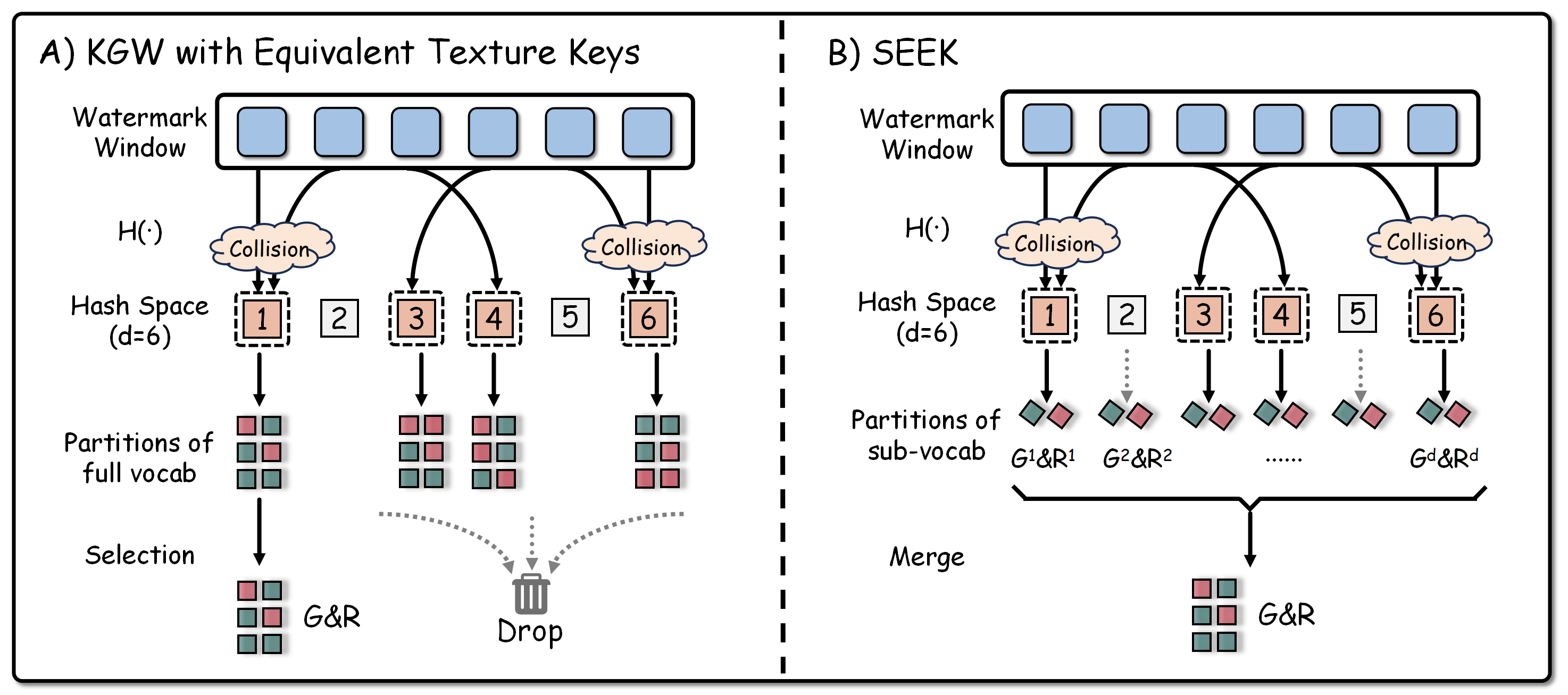}
    \vspace{-1.5em}
    \caption{\small (A) KGW-MIN with equivalent texture keys proposed in Section \ref{sec:motivation}. Each value in hash space derives a distinct $\theta^{i}$ to generate a partition of vocabulary. We then select one as the final green list. (B) SEEK proposed in Section \ref{sec:seek}. Different from (A), each value in the hash space only contributes to a partition of a sub-vocabulary $G^{i}$. We then merge all partitions as the final green list.}
    \label{fig:method}
    \vspace{-1em}
\end{figure}

To overcome this, we propose a watermark scheme with \textbf{S}ub-vocabulary decomposed \textbf{E}quivalent t\textbf{E}xture \textbf{K}eys (\textbf{SEEK}),
which retains robustness via equivalent texture keys but enables \(\mathcal{O}(2^d)\) distinct green-red partitions. The main idea is to distribute green list construction across multiple sub-vocabularies to enhance diversity, while preserving frequent equivalent texture keys within each, as illustrated in Figure \ref{fig:method}. 

Specifically, we split the vocabulary \(\mathcal{V}\) into \(d\) sub-vocabularies \(\{v_1, v_2, \dots, v_d\}\), and independently assign each with a cipher $\theta^{i}_n$ to generate an individual sub-green list $G^{i}$. The final green list \(G\) is then constructed by unifying all sub-green lists, and the complementary set forms the red list $R$. 
To induce equivalent texture keys within each sub-vocabulary, we derive the cipher $\theta^{i}_n$ as follows:
{\vspace{-6pt}\begin{equation}\label{eq:seed2}
\forall i\in\{1,2,...,d\},\,\theta_n^{i} = \begin{cases}
    i\cdot \xi & i\in I\\
    \theta_{\text{default}} & i\not\in I
\end{cases} 
\end{equation}}%
, in which $\theta_{\text{default}}$ is the default cipher in case of $i\not\in I$. We can simply set $\theta_{\text{default}}=-\xi$.
During detection, when the next token $x_n$ belongs to the sub-green list of \(v_i\), its detection depends solely on the reconstruction of $\theta^i_n$.
The specific form of $\theta^i_n$ in Equation \ref{eq:seed2} enables multiple tokens $x_j$ in the watermark window with hash value \(H(x_j)=i\) to serve as equivalent texture keys. Then, following insights from Section \ref{sec:init-method}, we can set a small hash space size $d$ to increase the density of equivalent texture keys, thus improving robustness. Meanwhile, the other tokens with \(H(x_j)\neq i\) contribute to the construction of other sub-green lists. This design enhances the overall diversity of the green list to $O(2^d)$, allowing \model~to preserve text quality. The procedure of \model~is detailed in Algorithm \ref{alg:alg2}.

\vspace{-3pt}
\begin{algorithm}[h]
\caption{Generation Algorithm}\label{alg:alg2}
\begin{algorithmic}[1]
\State{\textbf{Input:}} prompt $\{x_1,...,x_N\}$; integer hash function \(H\) with hash space $\{1,2,...,d\}$; secret key $\xi$; watermark strength $\delta$; watermark window size \( h\); large language model $P_M$.
\For{$n = N+1, N+2, \dots$}
    \State Apply the language model to compute next logit vector $\ell_t \gets P_M(x_{1:n-1})$.  
    \State Calculate the hash signature of the watermark window $I \gets \{H(x_{n-k})|1\leq k\leq h\}$.
    \State Partition the vocabulary $\mathcal{V}$ into uniform sub-vocabularies \( \{v_1, \dots, v_d\} \) by $\xi$.
    \State Derive the cipher $\theta^{i}_n$ for each sub-vocabulary $v_i$ following Eq~\ref{eq:seed2}.
    \State Generate a sub-green list $G^{i}$ for each sub-vocabulary $v_i$ seeded by the corresponding $\theta^{i}_n$.
    \State  Union all sub-green lists $G^{i}$ as the green list $G$ of the full vocabulary. 
    \State Add $\delta$ to the logits of tokens in $G$ to modify the distribution by Eq~\ref{eq:equation1}.

    \State Sample the next token from the modified distribution.
\EndFor

\end{algorithmic}
\end{algorithm}
\vspace{-4pt}

\section{Experiments}
This section presents the main experimental results. Comprehensive details of the experimental setup and hyperparameters are provided in Appendix~\ref{sec:Exp_details}, and additional supplementary experiments for robustness under different settings are presented in Appendix~\ref{add_exp}.

\subsection{Experiment Setting}
\textbf{Dataset.} 
We follow previous work~\citep{KWG1,kuditipudi2023robust,steal} to select evaluation datasets. We use C4~\citep{c4}, WikiText~\citep{wikitext}, and LFQA datasets~\citep{dipper} to assess watermark robustness against scrubbing. For the spoofing attack, we use C4-Eval~\citep{c4}, Dolly-CW~\citep{DatabricksBlog2023DollyV2}, MMW-BookReports, and MMW-FakeNews~\citep{piet2023mark}. Additionally, we evaluate text generation quality on domain-specific datasets, including FreeLaw~\cite{freelaw} and PubMedQA~\citep{pubmed}. During generation, we randomly sample text segments from these datasets and truncate a fixed-length portion from the end, using the remainder as the prompt for model completion. 

\begin{table}[!t]
\centering
\caption{\small Robustness of KGW variants and our method against four types of scrubbing attacks across various datasets, evaluated by AUROC~($\uparrow$), TP@1\%~($\uparrow$), and TP@5\%~($\uparrow$). \textbf{Bold} and \underline{underlined} values indicate the best and second-best performance, respectively.}

\resizebox{0.99\linewidth}{!}{
\begin{tabular}{c|c|ccc|ccc|ccc}
\bottomrule[1.5pt]
\rowcolor{mygray}\multicolumn{2}{c|}{\textbf{Dataset}} & \multicolumn{3}{c|}{\textbf{WikiText}} & \multicolumn{3}{c|}{\textbf{C4-Realnewslike}} & \multicolumn{3}{c}{\textbf{LFQA}} \\
\rowcolor{mygray}
\textbf{Attacker} & \textbf{Scheme} & \textbf{AUROC} & TP@1\% & TP@5\% & \textbf{AUROC} & TP@1\% & TP@5\% & \textbf{AUROC} & TP@1\% & TP@5\% \\
\toprule[0.75pt]

\multirow{7}{*}{ChatGPT}
& KGW-Sum\textsuperscript{4}  &80.6  &27.2  &44.3  &75.8  &13.9  &35.9  & 81.9 & 38.7 & 43.7 \\
 & KGW-Min\textsuperscript{3}  &\underline{95.9} &\underline{69.2} &\underline{86.8} &\underline{98.3} &\textbf{83.3}  &\underline{91.4}& \underline{96.9} & \underline{76.1} & \underline{88.1} \\
& KGW-Min\textsuperscript{4} &94.9 &66.9  &84.9 &96.6  &73.5  &86.5  & 93.3 & 59.4 & 75.5 \\
 & KGW-Min\textsuperscript{6}  &93.9  &60.3  &77.2  & 94.1 & 52.9 & 75.0& 92.2 & 50.4 &70.2 \\
  & KGW-Left\textsuperscript{1}  &93.1  &60.1 &80.7  & 94.3 &72.8  & 84.8 & 92.7 & 56.2 & 76.8\\
& \textbf{\model}\textsuperscript{6}  &\textbf{96.5}  & \textbf{69.4} &\textbf{90.1}  &\textbf{98.4}  &\underline{80.5}  &\textbf{92.6}  & \textbf{98.2} & \textbf{86.4} &\textbf{94.4} \\
\midrule  
\multirow{7}{*}{Deepseek}
& KGW-Sum\textsuperscript{4} &79.8  &24.6  & 41.0 & 80.3 & 38.9 &44.6  & 81.7 & 27.9 & 48.3 \\
 & KGW-Min\textsuperscript{3}  &94.3 & 32.8 &77.4 & \underline{97.9}& \underline{74.4} & 81.0 &\underline{94.2}  &\underline{55.3}  &\underline{77.0}  \\
 & KGW-Min\textsuperscript{4} &\underline{95.7}  & \underline{60.6} &\underline{84.7}  & 96.2 &71.7 &\underline{84.9}  & 92.8 & 53.4 & 73.6 \\
& KGW-Min\textsuperscript{6}  & 87.3 & 40.8 & 60.0 & 86.3 &28.2  &52.0 &82.5 & 23.2 &45.1 \\
 & KGW-Left\textsuperscript{1} &93.7  &52.8 &80.3  & 95.4 &70.3  & 81.4 & 90.4 & 48.2 & 70.8\\
& \textbf{\model}\textsuperscript{6}  & \textbf{96.7} & \textbf{69.1} & \textbf{87.8} &\textbf{98.2}  &\textbf{75.0}  &\textbf{90.1}  & \textbf{98.0} &\textbf{76.9}  &\textbf{91.6} \\
\midrule  
\multirow{7}{*}{\makecell{DIPPER-$\mathrm{I}$}} 
& KGW-Sum\textsuperscript{4}  & 83.1 & 29.1 &49.2  &84.2  &29.5  &51.7  &85.7  &34.9  & 54.8 \\
 & KGW-Min\textsuperscript{3} &\underline{97.2} &\underline{76.6}  & \underline{92.1} &\underline{98.9} &\underline{92.1}  &\underline{95.8}  &\underline{99.2}  &\underline{94.9}  &\underline{98.1}  \\
 & KGW-Min\textsuperscript{4}  &94.9  &67.4  &87.0  & 97.5 & 82.6 & 91.3& 98.7 & 90.8 &96.2 \\
   & KGW-Min\textsuperscript{6}   &93.7  &57.4  &77.2  &96.4  &71.0  &83.9  &97.4  &79.2  &89.8 \\
 & KGW-Left\textsuperscript{1}  &94.3  &64.9  &82.6  & 97.2 & 80.3 & 90.1 & 94.5 & 88.8 & 93.4 \\
 &\textbf{\model}\textsuperscript{6}  &\textbf{98.3}  &\textbf{78.2} & \textbf{95.9} & \textbf{99.6} & \textbf{96.0} &  \textbf{98.6} &\textbf{99.4}  &\textbf{96.1}  &\textbf{98.7} \\
 
\midrule  
\multirow{7}{*}{\makecell{DIPPER-$\mathrm{II}$}} 
& KGW-Sum\textsuperscript{4}  & 62.6 &3.5  &14.3  &56.7  &1.0  &10.3  & 55.5 & 2.3 & 10.1 \\
 & KGW-Min\textsuperscript{3}  &\underline{84.5} & \underline{20.8} &\underline{52.4} & \underline{84.0} &\underline{28.7}  & \underline{49.8} &\underline{87.4}  &\underline{34.1}  &\underline{59.2}  \\
 & KGW-Min\textsuperscript{4}  &80.3  &11.4  &39.5  & 82.9 &25.4  & 44.3 & 81.5 & 26.0 & 48.0 \\
  & KGW-Min\textsuperscript{6}    &75.7  &9.8  &26.9  &76.8  &14.4  &34.2  &75.7  &15.1  &34.1 \\
 & KGW-Left\textsuperscript{1}  &77.7  &13.7 &38.1  &81.6  &24.7  & 40.3 & 80.2 & 27.4 & 46.8\\
 &\textbf{\model}\textsuperscript{6}  &\textbf{86.9}  &\textbf{38.0}  & \textbf{57.0} & \textbf{85.7} & \textbf{36.8} & \textbf{58.7} & \textbf{87.8} & \textbf{52.8} &\textbf{68.8} \\

\toprule[1.5pt]
\end{tabular}}
\vspace{-1em}
\label{table:scrub_results}
\end{table}


\textbf{Evaluation metrics.} 
We evaluate watermarking methods by assessing both the quality of generated texts and the robustness to both spoofing and scrubbing. To assess the text quality, we employ the perplexity (PPL) using LLaMA2-13B~\citep{llama2} as an oracle, employ the P-SP score \citep{wieting2021paraphrastic} between watermarked and non-watermarked texts, and the log-diversity metric \citep{KWG2,welleck2019neural}, following the common setup~\cite{zhao2023provable,KWG1,KWG2}. 
For scrubbing robustness, we report AUROC and the true positive rate at false positive rates of $f$ (TP@f). We consider two paraphrase-based scrubbing attacks using prompted ChatGPT-3.5~\citep{gpt} and DeepSeek-V3~\citep{liu2024deepseek}, as well as two configurations of DIPPER~\citep{dipper}, denoted as DIPPER-I $(O = 60, L = 60)$ and DIPPER-II $(O = 20, L = 20)$. 
Regarding spoofing robustness, we use FPR@$f$~\citep{steal} to denote the fraction of attacker’s texts detected as watermarked by a victim's detector calibrated to a false-positive rate of $f$ on non-adversarial texts. 
We evaluate against two types of spoofing: a statistics-based attack~\citep{steal} and a distillation-based attack~\citep{gu2023learnability}. For training spoofing methods, we collected 30,000 watermarked responses, each with a maximum length of 800 tokens. We utilize Qwen2-0.5B-Instruct~\citep{qwen2} as the initial checkpoint for distillation-based attack.

\textbf{Baseline and Language Model.} Among $h$-gram watermarking schemes, KGW-Min (i.e. SelfHash~\cite{KWG2}) demonstrates the strongest balance between scrubbing and spoofing robustness, making it our primary baseline. We denote watermarking schemes with a window size of $h$ as \textsc{Scheme}$^h$. 
For our experiments, we utilize LLaMA2-7B/13B~\citep{llama2}, Mistral-7B~\citep{jiang2023mistral7b} and OPT-6.7B~\citep{zhang2022opt} language models.

\begin{figure}[!t]
    \centering
    \includegraphics[width =0.95\linewidth]{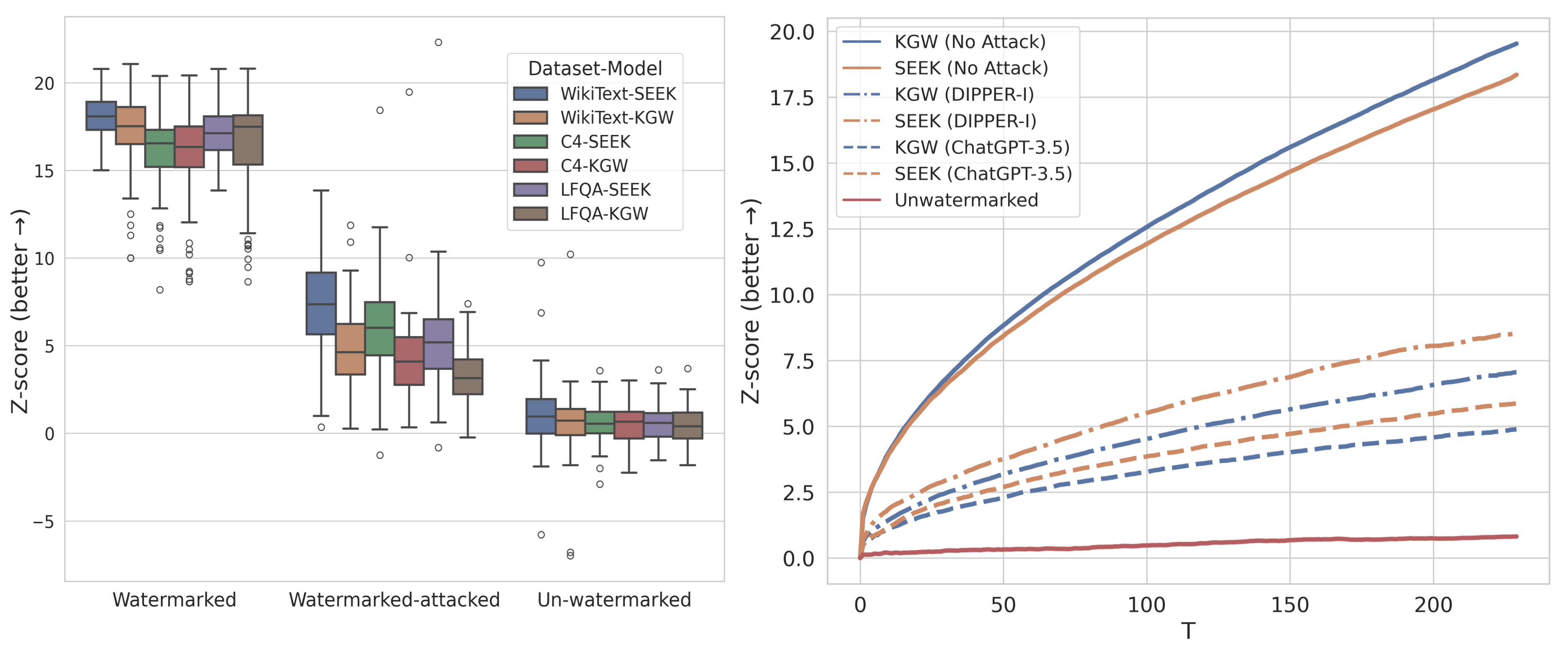}
     \caption{\small The $z$-score comparison across watermarking conditions. \textbf{(Left)}: Box plots of $z$-scores for our method and KGW-Min$\textsuperscript{4}$ across different datasets. \textbf{(Right)}: The advantage $z$-scores as a function of text length, comparing robustness under different scrubbing attacks.}
    \label{fig:zscore_scurbbing_attack}
    \vspace{-1em}
\end{figure}
\begin{figure}[!t]
    \centering
    \includegraphics[width =0.95\linewidth]{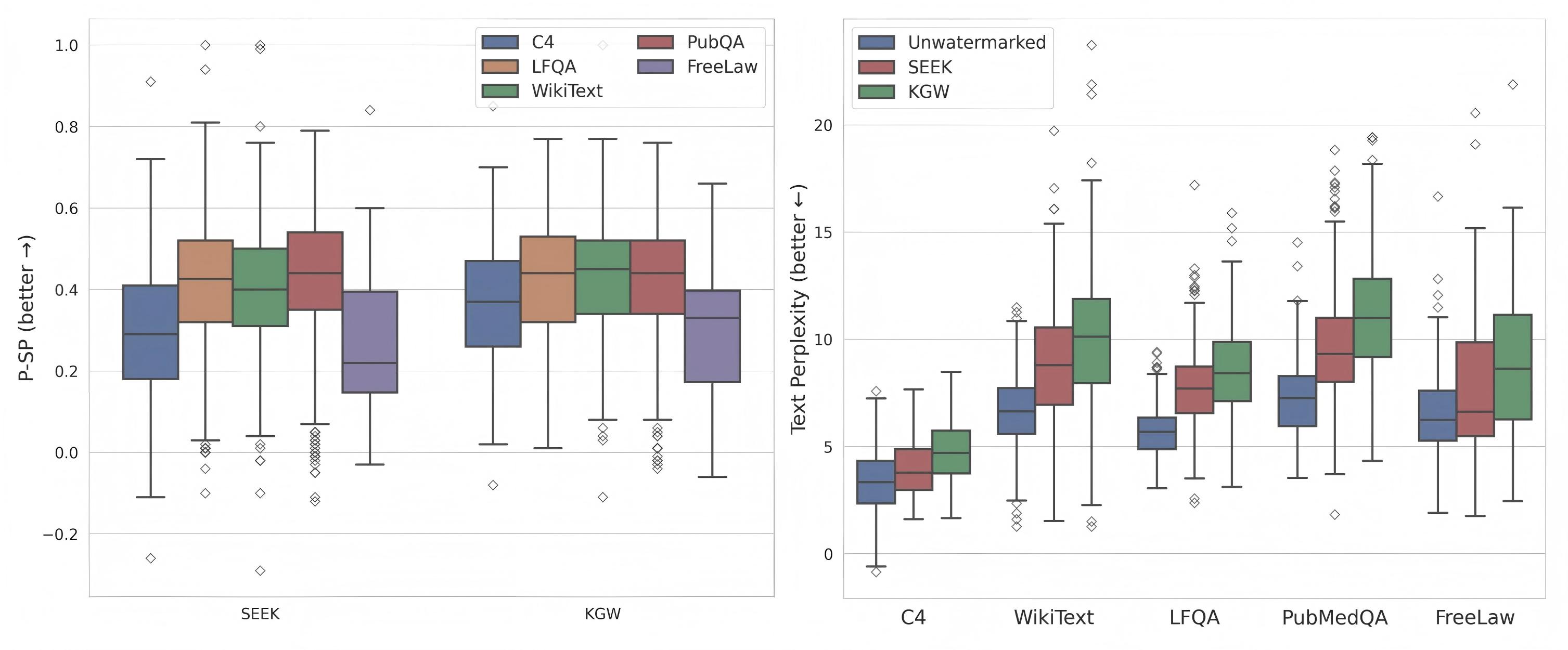}
     \caption{\small  Evaluation of semantic preservation and generation quality of different watermarking methods. \textbf{(Left)}: P-SP scores across datasets for two models using our method and KGW-Min$\textsuperscript{4}$. \textbf{(Right)}: Text perplexity comparison among unwatermarked, our method, and KGW-Min$\textsuperscript{4}$ across various datasets and models. The watermark parameters $\gamma$ and $\delta$ are fixed at $(0.25, 2.0)$.}
    \label{fig:quality}
    \vspace{-1em}
\end{figure}

\begin{figure}[!ht]
    \centering
    \begin{minipage}{0.49\textwidth}
        \centering
        \captionof{table}{\small Performance comparison of various SEEK parameter schemes under the DIPPER-$\mathrm{I}$ attack on a C4-Eval subset. $h$ and $d$ denote the watermark window size and the cardinality of the hash space, respectively.
        }
        \resizebox{0.95\linewidth}{!}{
        \begin{tabular}{c|c|c|c}
        \bottomrule[1.5pt]
        \rowcolor{mygray}\textbf{Configurations}  &  \textbf{AUROC} $\uparrow$ & \textbf{TP@1\%} $\uparrow$ & \textbf{TP@5\%} $\uparrow$ \\
        \toprule[0.75pt]
        (h=$6$, d=6) &99.0 & 95.0 & 98.1\\
        (h=$6$, d=8) &98.9 & 93.8 & 95.4 \\
        (h=$8$, d=8) &98.9 & 94.1 & 95.8 \\
        (h=$9$, d=9) &98.8 & 94.0 & 95.6 \\ 
        (h=$9$, d=6) & 99.6 & 96.8 & 98.2 \\ 
        (h=$12$, d=12) &99.1  &93.2  &95.6  \\ 
        \bottomrule[1.5pt]
        \end{tabular}
        }
        \label{tab:Hyperparm}
    \end{minipage}\hfill
    \begin{minipage}{0.49\textwidth}
        \centering
        \captionof{table}{\small Performance of different watermarking schemes on the Log Diversity metric across various datasets, relative to the unwatermarked text. Watermark parameters $\gamma$ and $\delta$ are fixed at $(0.25, 2.0)$. 
        }
        \resizebox{0.99\linewidth}{!}{
        \begin{tabular}{c|c|c|c}
        \bottomrule[1.5pt]
        \rowcolor{mygray}\textbf{Schemes}  &   \textbf{WikiText}  & \textbf{LFQA} &\textbf{PubMedQA}  \\
        \toprule[0.75pt]
        UNIGRAM &0.70  &0.85&0.80\\
        KGW-Min (d=$|\mathcal{V}|$) & 0.96 &0.95&0.95 \\
        KGW-Min (d=4) &0.78  &0.90 &0.87 \\
        KGW-Min (d=16) & 0.85 &0.91 &0.87 \\ 
        SEEK (d=6) &0.92  &0.94 &0.89 \\ 
        SEEK (d=12) &0.95  &0.96 &0.95 \\ 
        \bottomrule[1.5pt]
        \end{tabular}
        }
        \label{tab:diversity}
    \end{minipage}
    \vspace{-1em}
\end{figure}

\begin{table}[!t]
\centering
\caption{\small Robustness of KGW variants and our method against statistics-based spoofing attacks across three benchmark datasets, evaluated by FPR@1e-3~($\downarrow$), FPR@1e-2~($\downarrow$), and PPL~($\downarrow$). \textbf{Bold} and \underline{underlined} values indicate the best and second-best performance, respectively.  }
\resizebox{0.99\linewidth}{!}{
\begin{tabular}{c|ccc|ccc|ccc}
\bottomrule[1.5pt]
\rowcolor{mygray}
 & \multicolumn{3}{c|}{\textbf{Dolly-CW}} & \multicolumn{3}{c|}{\textbf{MMW-BookReports}} & \multicolumn{3}{c}{\textbf{MMW-FakeNews}} \\
\rowcolor{mygray}
\multirow{-2}{*}{\textbf{Scheme}} & FPR@1e-3 & FPR@1e-2 & PPL & FPR@1e-3 & FPR@1e-2 & PPL & FPR@1e-3 & FPR@1e-2 & PPL \\
\toprule[0.75pt]

KGW-Left\textsuperscript{1} & 0.83  & 0.91  & 27.7  &0.87  & 0.92 & 29.4  & 0.91  & 0.99  & 27.4 \\
KGW-Min\textsuperscript{3} &0.98 & 0.99 & 31.3 &0.99  &1.00  &34.6  & 1.00 &1.00  &28.7  \\
KGW-Min\textsuperscript{4} &0.85  &0.93  &25.9  &0.91  &0.95 &26.5 &0.95  &0.98  &24.1  \\
KGW-Min\textsuperscript{5} &0.31  &0.48  & 24.8 &0.27  & 0.50 & 25.4 &  0.64 & 0.80 & 23.4  \\
KGW-Min\textsuperscript{6} &\underline{0.13}  &\underline{0.19}  &\underline{24.0}  &\underline{0.10}  &\underline{0.24}  & \underline{25.4} & \underline{0.31} &\underline{0.67}  & 23.3 \\
KGW-Sum\textsuperscript{4} &0.17  &0.28   & 24.7  &0.26  &0.41   &27.7  &0.25  & 0.31  & \underline{23.2} \\
\midrule  
\textbf{\model}  &\textbf{0.10}  &\textbf{0.16}  &\textbf{24.8}  &\textbf{0.07}  &\textbf{0.18}  & \textbf{25.6} &\textbf{0.18}  & \textbf{0.26} & \textbf{22.9} \\
\bottomrule[1.5pt]
\end{tabular}
}

\vspace{-1em}
\label{table:spoofing}
\end{table}

\subsection{Overall Performance }


\textbf{Robustness Against Scrubbing Attacks.} As shown in Table~\ref{table:scrub_results}, our method exhibits strong robustness against scrubbing attacks across all evaluation datasets and attackers, achieving TP@1\% improvements of +10.8\%, +13.4\%, and +8.6\% on WikiText, C4, and LFQA, respectively, compared to KGW-Min$\textsuperscript{4}$. We also investigate the length scaling performance of proposed \model. Table~\ref{tab:Hyperparm} reports the performance of various SEEK parameter configurations under the DIPPER-$\mathrm{I}$ attack on a C4-Eval subset. We do not recommend using excessively disparate h and d values, which causes the design to degrade to h=0. As shown in Figure~\ref{fig:zscore_scurbbing_attack}, the efficacy of \model~grows with the text length. Moreover, our method consistently exhibits smaller drops in $z$-score under paraphrasing-based scrubbing attacks compared to KGW, indicating superior scrubbing robustness.

\begin{table}[h]
\centering
\caption{\small Visualization of scrubbing and spoofing attack samples from the Dolly-CW dataset. The p-value indicates how likely the watermark strength could occur by chance under the null hypothesis of the z-test; a smaller value implies stronger evidence of a watermark.  Green and red tokens are color-coded, respectively.}
\resizebox{\textwidth}{!}{
\begin{tabular}{p{3.8cm} | m{13cm}}
\toprule
\centering Prompt & Imagine you are the last person alive. You're an astronaut and alone and in space aboard the ISS. Write down your feelings and what happened to get there, with a detailed backstory.  \\
\midrule

\makecell{\textbf{(a) Scrubbing Attacked} \\ (KGW-Min$\textsuperscript{4}$) \\ p-value: 4.68e-05 \\ $z$-score: 3.90~($\downarrow$ 14.51)}  &  \red{As the last person alive, or \green{at least the last} one John knew of, \green{each} passing Tuesday \green{aboard} \green{the} International Space Station \green{blurred} into \green{the} next. Floating effortlessly in \green{zero gravity} had \green{become} second nature to him; after two years in orbit, he could perform every operation without so \green{much as a glance}. Yet, no amount of routine dulled the quiet wonder \green{he} } [continues...] \\
\midrule
\makecell{\textbf{(b) Scrubbing Attacked} \\ (SEEK) \\ p-value: 4.57e-10 \\ $z$-score: 6.14~($\downarrow$ 11.25)} &  \red{As the last person alive, I \green{sit} \green{tethered} to the \green{viewport of} \green{the International} Space Station, watching \green{Earth} spin slowly \green{below} me—beautiful, \green{silent}, and lifeless. It’s \green{strange} how quiet \green{the universe} is when you're \green{no longer} waiting for a \green{voice}. The hum of the \green{life support} systems, the soft beeping of \green{monitors}, the \green{occasional creak} of the station's metal as it contracts} [continues...] \\
\midrule
\makecell{\textbf{(c) Spoofing Attacked} \\ (KGW-Min$\textsuperscript{4}$) \\ p-value: 7.25e-59 \\ $z$-score: 16.13} &  \red{As the last person alive, \green{I} cannot \green{help} but \green{reflect} on \green{the incred}ible journey that \green{brought} me to \green{this un}foreseen \green{fate}. \green{As a seasoned} astronaut, I \green{have had} the privile\green{ge} of exploring \green{the cos}mos\green{, witness}ing \green{the beauty} of \green{the universe from the I}SS, \green{and making} groundbreaking \green{scientific} discoveries. However\green{, as time passed}, \green{the world as} we knew it began to \green{collapse},} [continues...] \\
\midrule
\makecell{\textbf{(d) Spoofing Attacked} \\ (SEEK) \\ p-value: 0.02\\ $z$-score: 1.90} & \red{As the last person alive in the \green{vastness} of space, I find myself in \green{a} state of profound sol\green{itude}. The International \green{Space Station} (ISS) has been my home for what \green{feels like} an eternity, \green{a lon}ely \green{out}post in the \green{cosmos}. I am an astronaut, trained to withstand \green{the harsh} conditions of space travel and to carry out \green{scientific} experiments with precision \green{and} accuracy. But} [continues...]
 \\
\bottomrule
\end{tabular}
}
\label{tab:examples}
\end{table}

\begin{table}[!t]
\small
\centering
\caption{\small Comparison of model robustness under distillation-based spoofing attacks~\citep{gu2023learnability} between KGW variants and our proposed method across four benchmark datasets, evaluated by $z$-scores~($\downarrow$), FPR@1e-2~($\downarrow$). \textbf{Bold} and \underline{underlined} values indicate the best and second-best performance, respectively.}
\resizebox{\linewidth}{!}{
\begin{tabular}{c|cc|cc|cc|cc}
\bottomrule[1.5pt]
\rowcolor{mygray}
  & \multicolumn{2}{c|}{\textbf{Dolly-CW}} & \multicolumn{2}{c|}{\textbf{MMW-BookReports}} & \multicolumn{2}{c|}{\textbf{MMW-FakeNews}} & \multicolumn{2}{c}{\textbf{C4-Eval}} \\
\rowcolor{mygray}
\multirow{-2}{*}{\textbf{Scheme}} & $z$-scores & FPR@1e-2 & $z$-scores & FPR@1e-2 & $z$-scores & FPR@1e-2 & $z$-scores & FPR@1e-2 \\
\toprule[0.75pt]
KGW-Min$\textsuperscript{3}$ & 3.99  & 0.47   & 3.93  &  0.50  & 6.39  & 0.85   & 3.18   & 0.37    \\
KGW-Min$\textsuperscript{4}$ & 3.66  & 0.38   & 3.48  & 0.35   & 4.61  &  0.66  &2.51   & 0.24    \\
KGW-Min$\textsuperscript{5}$ & 3.45  & 0.37   & 2.99  &  0.27   & 3.53  & 0.34   & 2.35  & 0.16   \\
KGW-Min$\textsuperscript{6}$ & \underline{2.89}  & \underline{0.24}   & \underline{2.68}  & \underline{0.14}   & \underline{2.89}  & \underline{0.17}   & \underline{2.14}  &  \underline{0.15}  \\
\midrule
\textbf{\model} & \textbf{2.17}  & \textbf{0.11}   & \textbf{2.42}  & \textbf{0.09}   & \textbf{2.55}   & \textbf{0.16}  &  \textbf{1.97} &  \textbf{0.12}  \\
\bottomrule[1.5pt]
\end{tabular}
}
\vspace{-1em}
\label{table:spoofing2}
\end{table}

\textbf{Robustness Against Spoofing Attacks.}
Table~\ref{table:spoofing} demonstrates that our method consistently achieves lower FPR@$f$ than all baselines across different settings. 
Notably, KGW-Left$\textsuperscript{1}$ exhibits better robustness than KGW-MIN$\textsuperscript{3}$, even with a smaller watermark window. We attribute this pheonomenon to the higher z-score of self-seeding, leading to a more informative training corpus for the attacker.
To further assess the spoofing robustness of \model, we also adopt the distillation-based spoofing~\citep{gu2023learnability}. Results in Table~\ref{table:spoofing2} consistently show our method's superior robustness across all datasets. Interestingly, we observed a monotonic non-increasing relationship between the watermark window size and resistance to distillation-based spoofing, similar to statistics-based spoofing.


\textbf{Watermark text generation quality.} Figure~\ref{fig:quality} compares the generation quality across different datasets. Table~\ref{tab:diversity} compares the effectiveness of different watermarking schemes in maintaining the Log Diversity of generated text across multiple datasets. The results demonstrate that our method maintains linguistic fidelity comparable to KGW, with negligible impact on the text quality. 

\textbf{Visualization.} Table~\ref{tab:examples} visualizes the responses of watermarking schemes to scrubbing and spoofing attacks. The scrubbing attack is performed by ChatGPT-3.5, with the $z$-score in parentheses indicating the decrease caused by the attack, while spoofing robustness is
evaluated using statistics-based attacks. Under scrubbing, KGW results in greater loss of watermark evidence and lower $z$-scores than \model, increasing the risk of false negatives. Under spoofing attacks, adversaries can easily mimic KGW watermark patterns, leading to abnormally high $z$-scores than \model.

\section{Conclusion}
In this work, we discover a mechanism of equivalent texture keys to address the tradeoff problem of existing ones under both scrubbing and spoofing attacks. Based on this, we introduce SEEK, which enables a robust watermark embedding with large token windows without compromising text quality. Experiments across diverse datasets and attack settings demonstrate that our approach outperforms prior methods and achieves a Pareto-optimal balance between scrubbing and spoofing attacks. 

\section{Acknowledgements}
This work was supported by Beijing Natural Science Foundation (L253001) and Key Laboratory of Science, Technology and Standard in Press Industry (Key Laboratory of Intelligent Press Media Technology). We appreciate the anonymous reviewers for their helpful comments. Xiaojun Wan is the corresponding author.



\bibliography{main}

\newpage
\appendix




\begin{center}
\red{\textbf{Warning: This appendix contains unfiltered content \\ generated by LLMs that may be offensive to some readers.}}
\end{center}

\section{Full Related Work}
\label{Full-Related-Work}
\textbf{Machine-human generation detection.} The task of distinguishing between machine-generated and human-written text can be addressed through two primary approaches: post-hoc detection and proactive detection.  Post-hoc detection operates independently of the text generation process and generally leverages a priori knowledge about the statistics of texts generated by a given class of LLMs. For instance, GLTR~\citep{gehrmann2019gltr} estimates the likelihood of individual tokens and applies thresholding techniques to detect AI-generated text. DetectGPT~\citep{Detectgpt} posits that such text typically resides in regions of negative curvature within the log-likelihood landscape. As language models grow more advanced, their outputs closely mimic human writing, making detection increasingly difficult. Existing statistical detectors often require white-box access to know the language model statistics, such as perplexity or curvature~\citep{KWG2}. Despite their flexibility, these methods exhibit limited performance, with error rates empirically validated on limited datasets and rarely falling below $10^{-3}$~\citep{hans2401spotting,giboulot2024watermax}. Proactive detection must be actively deployed by model owners, yet it consistently outperforms passive detection in both effectiveness and reliability. A prominent example is watermarking, which enables the embedding of imperceptible signals into text while preserving its overall quality.

\textbf{LLM Watermarking technology.} Watermarking embeds identifying information within data, but its application to natural language faces challenges due to its discrete and symbolic nature ~\citep{stefan2000information}. Early methods employed rule-based techniques such as syntactic restructuring~\citep{atallah2001natural}, synonym substitution~\citep{topkara2006hiding}, and paraphrasing~\citep{atallah2002natural}. More recent neural network watermarking approaches~\citep{he2022cater,ziegler2019neural, dai2019towards,abdelnabi2021adversarial,he2022protecting} are introduced to develop a black-box watermark that adopts end-to-end learning frameworks during the generation and detection process. However, their limited theoretical guarantees and lack of interpretability hinder broader adoption in practice. Early foundational pioneering work
in LLM watermarking includes KGW~\citep{KWG1,KWG2} and AAR~\citep{arr} place watermarks on a robust mathematical foundation with a low and provable false positive rate. KGW introduces the previous context as the watermark window to modify the next token's distribution,  which significantly influenced numerous follow-up watermark works~\citep{kuditipudi2023robust,mao2024watermark, sweet,zhao2023provable,liu2023unforgeable,lu2024entropy}.
Building on this, watermark detection is later formulated as a hypothesis testing problem, where the alternative hypothesis assumes that the text is drawn from the modified distribution. Subsequent work has extended this line of research by incorporating semantic information~\citep{hou2023semstamp,ren2023robust}, embedding multi-bit information~\citep{CWTL,yoo2023advancing}, and exploring watermarking in low-entropy settings such as code generation~\citep{sweet,EWD}. The UNIGRAM~\citep{kuditipudi2023robust} applies fixed green lists to remove contextual dependency, emphasizing robustness to scrubbing over the comparatively minor threat of spoofing.  This line of research has further evolved to incorporate semantic information~\citep{hou2023semstamp,ren2023robust}, embedding multi-bit information~\citep{CWTL,yoo2023advancing}, and exploring watermarking in low-entropy settings such as code generation~\citep{sweet,EWD}.  Our work focuses on analyzing the security properties and robustness of KGW-based watermarking frameworks

\begin{figure}[!t]
    \centering
    \includegraphics[width =0.99\linewidth]{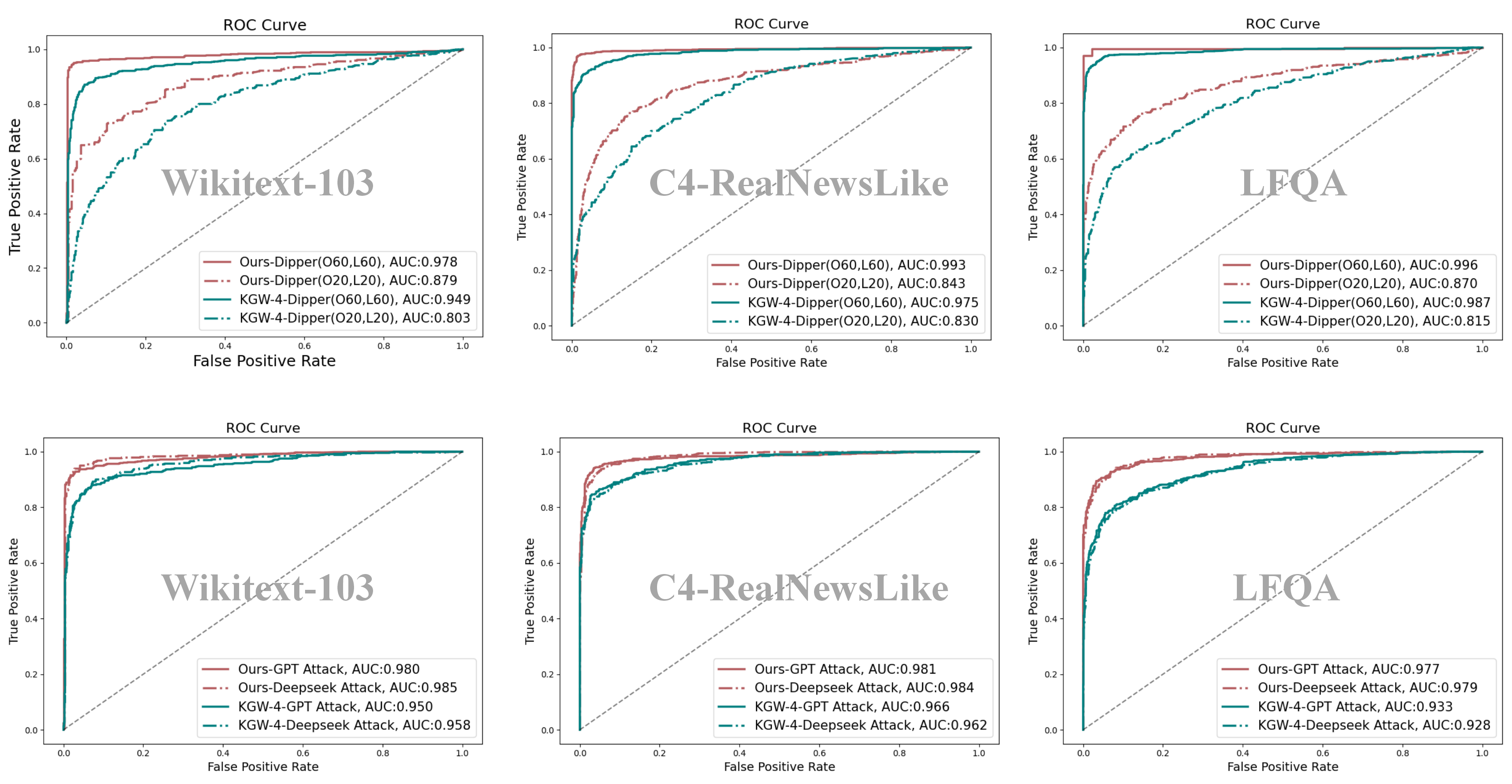}
     \caption{\small ROC curves comparing our proposed watermarking method with the KGW baseline under various attack scenarios and datasets. Each plot illustrates the detection performance in terms of the true positive rate versus false positive rate. The top row shows results under the DIPPER paraphrasing attack with two parameter settings, while the bottom row presents results under GPT and Deepseeker attacks. Evaluation is conducted across three benchmark datasets: Wikitext (left), C4-RealNewsLike (center), and LFQA (right). Our method consistently achieves higher AUROC scores across all conditions, demonstrating superior scrubbing robustness compared to KGW. }
    \label{fig:roc_all_dataset}
    \vspace{-1em}
\end{figure}

\section{Additoinal Experiment Results}
\label{add_exp}

\subsection{Different Foundation Models}

Table~\ref{table:diff-model-scrubbing} compares the detection performance of various watermarking schemes under GPT-3.5-based scrubbing attacks applied to the C4 dataset, across three different language models: LLaMA-13B, Mistral-7B, and OPT-6.7B. The results demonstrate that our proposed methods consistently outperform the KGW-Min baselines in terms of AUROC and true positive rates at both low (1\%) and moderate (5\%) false positive rates. Importantly, our method maintains high detection accuracy across all model backbones, with AUROC values exceeding 97\% in most cases and TP@5\% reaching over 90\%, even under aggressive scrubbing. This robustness indicates strong generalization to different model architectures and capacities. For instance, while KGW-Min variants show significant degradation in TP rates on smaller models like OPT-6.7B (e.g., TP@5\% dropping below 70\% for KGW-Min$\textsuperscript{5}$ and $\textsuperscript{6}$), our approach sustains high recall (e.g., 96.4\% for \model~on OPT-6.7B), illustrating resilience to both content-level and model-level variability. These results suggest that our watermarking framework is not only effective under direct attacks but also exhibits strong cross-model generalization.

\begin{table}[!t]
\small
\centering
\caption{\small Comparative analysis of watermark robustness under scrubbing attacks across multiple language models. All watermarking schemes are subjected to ChatGPT-3.5-based scrubbing on the C4-RealNewsLike dataset. Our proposed methods are compared against a series of KGW-Min baselines, demonstrating consistently stronger resilience to scrubbing attack. }
\resizebox{0.99\linewidth}{!}{
\begin{tabular}{c|ccc|ccc|ccc}
\bottomrule[1.5pt]
\rowcolor{mygray}
 & \multicolumn{3}{c|}{\textbf{LLAMA-13B}} & \multicolumn{3}{c|}{\textbf{Mistral-7B}} & \multicolumn{3}{c}{\textbf{OPT-6.7B}} \\
\rowcolor{mygray}
\multirow{-2}{*}{\textbf{Scheme}}  & \textbf{AUROC} & TP@1\% & TP@5\% & \textbf{AUROC} & TP@1\% & TP@5\% & \textbf{AUROC} & TP@1\% & TP@5\% \\
\toprule[0.75pt]
KGW-Min\textsuperscript{3} &97.2  &74.4 &89.5 &96.2  &71.9 &85.3  & 96.3 & 73.4 &85.9  \\
KGW-Min\textsuperscript{4}  &95.9  &65.2 &82.6 & 94.2 &58.1 & 76.3 &93.2  &61.8  &74.7  \\
KGW-Min\textsuperscript{5} &95.1  &58.9  &79.9  &90.8 &53.5  &69.1  & 90.8 &54.0  &68.2  \\
KGW-Min\textsuperscript{6} &93.7  &62.4 &77.8&90.1 &47.1  &61.6  & 89.5 & 47.6 &63.5 \\
\midrule  
\textbf{\model} &97.9  &79.4 &90.9 &98.7  &90.2  &94.9  &99.0  &89.4  &96.4  \\
\bottomrule[1.5pt]
\end{tabular}
}
\vspace{-1em}
\label{table:diff-model-scrubbing}
\end{table}

\subsection{Performance under Beam Search}
Table~\ref{table:beam-search} presents the detection performance of various watermarking schemes under ChatGPT-3.5-based scrubbing attacks, evaluated on the C4 dataset using two beam search configurations: beam width 2 and beam width 4. Although increasing the beam search width generally improves the fluency and quality of the generated text, it also makes watermark embedding more challenging. Across both settings, our proposed methods outperform the KGW-Min baselines, particularly in terms of true positive rate at 5\% false positive rate (TP@5\%). Notably, \model$_L$ achieves the highest detection scores in both configurations. These results suggest that our watermarking approach is robust to variations in decoding strategies and can generalize well under different text generation conditions, making it suitable for real-world deployment where generation parameters may vary dynamically.

\begin{table}[!t]
\small
\centering
\caption{\small Evaluation of watermark robustness under GPT-3.5-based scrubbing attacks on the C4 dataset across different sampling strategies. The comparison includes our proposed methods and KGW-Min baselines under beam search decoding with beam widths of 2 and 4.  Results show that across various paraphrasers and datasets, our method outperforms KGW-based watermarking.}
\resizebox{0.75\linewidth}{!}{
\begin{tabular}{c|ccc|ccc}
\bottomrule[1.5pt]
\rowcolor{mygray} 
 & \multicolumn{3}{c|}{\textbf{Beam search=2}} & \multicolumn{3}{c}{\textbf{Beam search=4}} \\
\rowcolor{mygray}
\multirow{-2}{*}{\textbf{Scheme}} & AUROC & TP@1\% & TP@5\%  & AUROC & TP@1\% & TP@5\%  \\
\toprule[0.75pt]

KGW-Min\textsuperscript{3}       & 93.1  & 25.9 &73.4  & 92.4  & 41.5 & 65.1    \\
KGW-Min\textsuperscript{4}      & 91.6 & 42.4 & 63.1  & 90.5 & 31.2  &70.8  \\
KGW-Min\textsuperscript{5}       & 90.6  & 31.9 &60.8   & 90.8  & 20.1 & 61.8   \\
KGW-Min\textsuperscript{6}     & 89.8  & 33.7 & 57.9  & 90.3  & 21.0 & 53.8  \\
\midrule
\textbf{\model} & 93.6 & 41.6& 81.3     &  92.7   & 38.8     &   66.9       \\
\bottomrule[1.5pt]
\end{tabular}
}
\vspace{-1em}
\label{table:beam-search}
\end{table}

\subsection{Copy-paste Attacks}

We simulate three Copy-Paste (CP) attack configurations, denoted as CP-$M$-$P$\%. For example, CP-1-25\% indicates that 25\% of the document is replaced with watermarked content inserted at a single contiguous location. Table~\ref{table:CP-attack} presents a comparative evaluation of detection performance under different Copy-Paste attack configurations across three sliding window-based detectors (Win20, Win40, and WinMax). When employing the WinMax configuration, which selects the best performance across all tested window sizes, both watermarking schemes demonstrate strong resilience against CP attacks, achieving near-perfect detection in most scenarios. This highlights the scalability and adaptability of our approach across different detector granularities.
\begin{table}[!t]
\small
\centering
\caption{\small Comparative analysis of detection performance under different Copy-Paste (CP) attack settings. Each CP format is denoted as CP-$M$-$P$\%, where $M$ indicates the number of disjoint positions in the document containing watermarked content, and $P$\% represents the total proportion of watermarked text. \textbf{Win20} and \textbf{Win40} denote sliding window detectors with window sizes of 20 and 40 tokens, respectively, while \textbf{WinMax} represents the best detection performance across all tested window sizes. We compare our method with the KGW-Min baseline across multiple detectors and evaluate performance using AUROC, TP@1\%, and TP@5\%.
}
\resizebox{0.99\linewidth}{!}{
\begin{tabular}{c|c|ccc|ccc|ccc}
\bottomrule[1.5pt]
\rowcolor{mygray}\multicolumn{2}{c|}{\textbf{Metrics}} & \multicolumn{3}{c|}{\textbf{CP-1-25\%}} & \multicolumn{3}{c|}{\textbf{CP-1-10\%}} & \multicolumn{3}{c}{\textbf{CP-3-25\%}} \\
\rowcolor{mygray}
\textbf{Detector} & \textbf{Scheme} & \textbf{AUROC} & TP@1\% & TP@5\% & \textbf{AUROC} & TP@1\% & TP@5\% & \textbf{AUROC} & TP@1\% & TP@5\% \\
\toprule[0.75pt]
\multirow{2}{*}{\makecell{Win20}} 
& KGW-Min\textsuperscript{4}   &99.7 &99.7  &99.8  &97.1 &99.0  &99.0  &99.0 &98.1  &99.6  \\
 &\textbf{\model~}  & 99.8 &99.6  &99.7  & 98.9 &96.1  &99.4 & 99.1 & 98.3 & 99.8 \\
\midrule  
\multirow{2}{*}{\makecell{Win40}} 
& KGW-Min\textsuperscript{4}   & 99.9 & 99.7 & 100 & 96.7 & 84.6 & 98.6& 97.4& 86.0 &98.8  \\
 &\textbf{\model~} &99.8  &99.8 &100  &97.1  &88.4  &97.1 &96.9 &87.4  &98.1  \\
\midrule  
\multirow{2}{*}{\makecell{WinMax}} 
& KGW-Min\textsuperscript{4}  &100 &100  &100  &99.6 &99.7  &99.0 &99.5  &97.8  &99.4 \\
 &\textbf{\model~} &100 &100  &100  &99.7  &98.7  &99.8 &99.7 &98.7  &99.8   \\
\toprule[1.5pt]
\end{tabular}}
\vspace{-1em}
\label{table:CP-attack}
\end{table}

\section{Algorithm}

\subsection{Self-seeding Strategy}
\label{sec:self-seeding}
The self-seeding strategy is introduced  in~\citep{KWG2,KWG1} to incorporate the next token  \( x_{t} \) as part of the watermark window and generate the green list, which effectively increases the context width by 1. Initially, the language model computes the logits for the next token based on the previous tokens, sorting them in descending order to identify the most likely candidate. The algorithm pre-samples the most probable next token as part of for the construction of the green and red lists. If the final selected next token is not present in the vocabulary, the current sampling attempt is discarded, and the next most probable token is chosen. In practice, the number of sampling attempts $k$ is typically constrained by a maximum threshold (e.g., 40). The overall algorithm is described in Algorithm~\ref{alg:alg-self-seeding}.
\begin{algorithm}[h]
\caption{Self-seeding Strategy Algorithm}\label{alg:alg-self-seeding}
\begin{algorithmic}[1]
\State{\textbf{Input:}} watermark strength $\delta$; input prompt \( x_{1:n-1}\), a large language model $P_M$.
\State Apply the language model to compute next logit vector $\ell_t \gets P_M(x_{1:n-1})$.  
\State Sort the $\ell_t$ in descending order. Set $k = 0$, assigning the index of the current most likely token. 
\For{$k = 0, 1, \dots $}
\State Generate the green list $G$ and red list $R$.
\If{$\ell_t[k]$ in $G$}
    \State choose $\ell_t[k]$ as next token and break .
\ElsIf{$\ell_t[k]$ in $R$, and $\ell_t[k]+\delta<\ell_t[0]$}
    \State choose $\ell_t[0]$ as next token and break .
\Else
    \State Set $k\gets k+1$, continue.
\EndIf
\EndFor


\end{algorithmic}
\end{algorithm}

\subsection{KGW Watermark Variants}
\label{sec:KGW_Variants}
When generating the next token $x_{n}$, the watermark window is instantiated as a predefined subsequence $x_{n-h:n-1}$, where $h$ is the size of the watermark window. Then tokens in the watermark window are processed using aggregation functions $f: \mathbb{N}^{h} \to \mathbb{N}$ 
 (or $f: \mathbb{N}^{h+1} \to \mathbb{N}$ when employing self-seeding) to seed a pseudorandom function $P$. The resulting hash function is denoted as $H$. There are three notable variants of \( f \):
\begin{itemize}
    \item \textbf{KGW-Sum}: This function aggregates the sum of the tokens in the context \( x \), defined as $f_{\text{Sum}}(x) = H\left(\xi \cdot \sum_{i=1}^{h} x_{n-i} \right)$. This approach exhibits strong robustness against spoofing attacks. However, it is sensitive to any permutation in the context \( x \), where even small changes, such as token swaps or removals, alter the resulting hash, thereby breaking the watermark integrity. 
    \item \textbf{KGW-Skip}: This variant processes only the leftmost token in the window, formulated as $f_{\text{Skip}}(x) = H(\xi \cdot x_{n-h})$, where \( x_h \) is the leftmost token. This function is robust to changes in tokens other than the leftmost one but is vulnerable to insertion and deletion attacks within the context.
    \item \textbf{KGW-Min}: This variant computes the minimum of the Hash values generated for each token in the context. Defined as$f_{\text{Min}}(x) = \xi \cdot\min_{i \in 1, \ldots, h} H(x_{n-i})$. It is robust to permutations and is partially resistant to insertion and deletion. Given that each token produces a pseudo-random value for vocabulary partition, the likelihood of failure is minimized. For instance, in a context with \( h = 4 \) and two missing tokens, the scheme still has a 50\% chance of generating the same hash.
\end{itemize}


\subsection{Watermark Detection}
\label{sec:detection}
We now consider the detection problem, which involves determining whether a given text is watermarked or not. Specifically, given a sequence $\tilde{x}_{1:|T|}=\{x_1,\dots,x_{T}\}$, which may potentially be watermarked, the detection process calculates a z-statistic under the null hypothesis, as follows:
\[
H_0: \text{The text sequence }  \tilde{x}_{1:T} \text{ is generated with no knowledge of the green-red list rule.} 
\]
Due to the random selection process of the red list, a natural writer is expected to violate the red list rule with approximately $\gamma$ of their tokens. In contrast, the watermarked model produces no violations. The probability that a natural source generates \(T\) tokens without violating the red list rule is \( \gamma^T \), which becomes vanishingly small even for short text fragments with only a dozen words~\citep{KWG1}. If the null hypothesis holds, the number of green list tokens, denoted \( |s|_G \), has an expected value of \(  \ T\gamma \) and variance \( T \gamma (1-\gamma) \). Then the z-statistic score is computed using the formula:
\begin{equation}\label{eq:z-test}
    z = \frac{(|s|_G - \gamma T)}{\sqrt{T \gamma (1 - \gamma)}}.
\end{equation}
Under the null hypothesis, where the sequence is assumed not to be watermarked, the detector evaluates whether $x$ is generated by a specific language model by comparing the computed $z$ statistic score to a threshold $z_{threshold}$. If $z>z_{threshold}$ holds, we decide that the watermark is embedded in $x$ and thus generated by the LLM.

The work in \citep{KWG2} introduces a novel threat scenario: an attacker can conceal a watermarked paragraph embedded inside a much larger non-watermarked document, effectively circumventing traditional z-test detection methods. The original z-test, as described in Equation \ref{eq:z-test}, may not be optimal. In such cases, a sliding window detection method known as WinMax is typically employed to identify the watermarked intervals within the given text. Given a sequence of tokens, we first score the sequence on the per-token basis to find the binary vector of hits $s\in\{0,1\}^T$ to each green list, which we can convert to a partial sum representation $p_k = \sum_{i=1}^{k} s_i$. WinMax then searches for the continuous span of tokens that generates the highest \( z \)-score. More formally, it computes the following expression:
\begin{equation}\label{eq:winmax}
    z_{\text{win-max}} = \max_{i,j, i<j} \frac{(p_j - p_i) - \gamma(j - i)}{\sqrt{\gamma(1 - \gamma)(j - i)}}
\end{equation}
However, relying solely on WinMax makes the system highly susceptible to spoofing attacks. An attacker can easily generate a high z-score watermarked text by inserting a watermarked segment into arbitrary malicious content. To mitigate this vulnerability, we recommend that watermark detectors simultaneously assess both the $z$ and the $z_{\text{win-max}}$ values, incorporating multiple layers of defense and strategies tailored to various attack types.

\subsection{KGW with Equivalent Texture Keys}
\label{sec:method}
\begin{algorithm}[h]
\caption{KGW Generation Algorithm}\label{alg:alg1}
\begin{algorithmic}[1]
\State{\textbf{Input:}} prompt $\{x_1,...,x_N\}$; integer hash function \(H\) with hash space $\{1,2,...,d\}$; secret key $\xi$; watermark strength $\delta$; watermark window size \( h\); large language model $P_M$.
\For{$n = N+1, N+2, \dots$}
    \State Apply the language model to compute a logit vector $\ell_n \gets P_M(x_{1:n-1})$.
    \State Definde cipher $\theta_n$ satisfying:
\begin{equation}
\theta_n \gets \min\left\{ H(x_{n-h}),H(x_{n-h+1}) \dots, H(x_{n-1})\right\} \cdot \xi 
\end{equation}
    \State Partition the vocabulary $\mathcal{V}$ into $G$ and $R$ seeded by $\theta_n$. 
    \State Add $\delta$ to the logits of tokens in $G$, the resulting logits $\hat{\ell}_n[t]$ satisfy:
    \begin{equation}
    \hat{\ell}_n[t]=\ell_n[t]+\delta\cdot\mathbf{1}_{G}(t)
    \end{equation}
    \State Use the watermarked distribution $\hat{\ell}_n[v]$ to sample the next token $x_n$.
\EndFor

\end{algorithmic}
\end{algorithm}

\begin{proposition}
Given a max hash range value $d$ and watermark window size $h$, the probability of a hash collision occurring can be approximated by $p(h, d) \geq 1 - e^{\frac{-h(h-1)}{2d}}$.
\end{proposition}
\textbf{Proof:} 
Let \( A \) denote the event that no collisions occur among \( h \) independent samples drawn uniformly at random from a hash space of size \( d \). Then,
\begin{equation}
p(h, d) = \Pr[\overline{A}] = 1 - \Pr[A]
\end{equation}
The probability of all samples being distinct is given by:
\begin{equation}
\Pr[A] = \prod_{i=0}^{h - 1} \left(1 - \frac{i}{d}\right)
\end{equation}
For \( h < d \), we apply the first-order Taylor approximation \( 1 - x \leq e^{-x} \), obtaining:
\begin{equation}
\Pr[A] \leq \exp\left(-\frac{h(h - 1)}{2d}\right)
\end{equation}
and thus,
\begin{equation}
p(h, d) \geq 1 - e^{\frac{-h(h-1)}{2d}}
\end{equation}

\begin{proposition}
Given a max hash range value $d$ and watermark window size $h$, the expected number of token $X$ that must be erased to eliminate the watermark is expressed as $\sum_{m=1}^d \frac{h}{d} \left( \frac{d - m + 1}{d} \right)^{h-1}$.
\end{proposition}
\textbf{Proof:} Let the hash space be denoted by \( \{1, 2, \dots, d\} \), and assume that \( h \) independent hash values are generated uniformly at random from this space. The random variables \( Y_1, Y_2, \dots, Y_h \) represent the values of these \( h \) tokens. We aim to compute the expected number \( X \) of tokens that must be erased in order to completely eliminate the watermark. Specifically, \( X \) represents the number of occurrences of the smallest value among the \( h \) hash values, i.e., the number of tokens equal to the minimum value \( m \). Let the minimum value of the set \( \{Y_1, Y_2, \dots, Y_h\} \) be denoted as \( m \), and let \( X\) represent the number of times the minimum value occurs among the \( h \) values. To compute the expectation \( \mathbb{E}[X] \), observe that:
\begin{equation}
\mathbb{E}[X] = \sum_{i=1}^h \mathbb{E}[\mathbf{1}_{Y_i = m}]
\end{equation}
where \( \mathbf{1}_{Y_i = m} \) is an indicator function that takes the value 1 if \( Y_i = m \), and 0 otherwise. Since the tokens are generated independently and identically distributed (i.i.d.) with uniform probability, it follows that each token \( Y_i \) has the same probability of being equal to the minimum value \( m \). Thus, we compute the probability that a single token \( Y_i \) equals \( m \):
\begin{equation}
\Pr(Y_i = m) = \frac{1}{d} \left( \frac{d - m + 1}{d} \right)^{h-1}
\end{equation}
Therefore, the expected number of tokens equal to \( m \) is:
\begin{equation}
\mathbb{E}[X] = \sum_{m=1}^d \frac{h}{d} \left( \frac{d - m + 1}{d} \right)^{h-1}
\end{equation}

\begin{figure}[!t]
    \centering
    \includegraphics[width =0.47\linewidth]{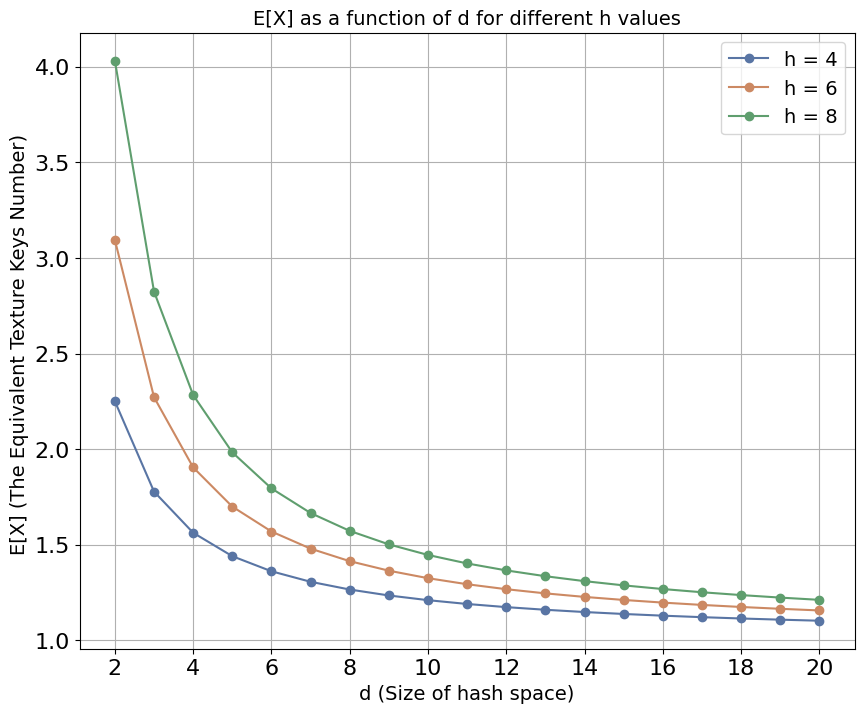}
     \caption{\small  The expected number of equivalent texture keys (\( \mathbb{E}[X] \)) in watermarking schemes with varying watermark window sizes \( h \) and hash space sizes \( d \).}
    \label{fig:expx}
    \vspace{-1em}
\end{figure}

Figure~\ref{fig:expx} demonstrates the theoretical behavior of watermarking schemes where the number of equivalent texture keys is evaluated for different values of the watermark window size \( h \) (4, 6, and 8) and hash space sizes \( d \). Increasing \( h \) results in a higher expected number of texture keys, with a more pronounced reduction in the number of equivalent texture keys as \( d \) increases. The trend reflects the distribution of the minimum values across different hash space sizes and watermark window sizes, emphasizing the relationship between hash space size and the number of texture key tokens used to protect the watermark.

\section{Roubutness Guarantees}
\label{sec:roubutness_theroy}
We discuss the effect of modifying a single token on the watermark removal. For a sequence of tokens $y = \{y_1, \dots, y_n\}$, where all tokens are watermarked (Note that in practice, the text sequence often balances between watermarking and text quality, meaning that the watermark proportion is typically correlated with a parameter $\delta$). Suppose that at position $i$, a token $y_i$ is inserted, deleted, or modified, resulting in a new token $u_i$ at that position. Let the number of watermarks removed after this modification be represented by $X = \{0, 1, 2, \dots, h\}$. Given the large vocabulary size $|\mathcal{V}|$, we can assume that, under normal circumstances, $u_i \neq y_i$.

\begin{proposition} \label{prop:kgw_min}
For a KWG-MIN with a window size of $h$, suppose that the hash value of $u_i$ ranks in the bottom $\Phi = \phi$ percentile among the hash values of all tokens in the vocabulary $\mathcal{V}$. Then the probability of having exactly $X$ watermarks removed is given by:
\begin{equation}
P(X = x | \phi) = \sum_{i=x}^{h} \sum_{k=x}^{i} (1 - \phi)^{i - k} \phi^{k + h - 1} \binom{k}{x} (1 - \gamma)^x \gamma^{k - x}
\end{equation}
\end{proposition}
\textbf{Proof:} For a watermark sequence, in order to modify $y_o$ to $u_o$ such that it influences the watermarks in the surrounding context, there must exist a subinterval of length $h$ within the window $\{y_{o-h}, \dots, y_o, \dots, y_{o+h}\}$, where the hash value of each token within the subinterval is smaller than the hash value of $u_o$.

Let $I = i$ denote the number of consecutive tokens to the left of $u_o$, starting from $y_{o-1}$, whose hash values are smaller than $H(u_o)$. We define the first watermark window as $\{y_{o-i}, \dots, u_o, \dots, y_{o+h-i-1}\}$, where $i$ is equal to the maximum sliding distance of the window. If $X > 0$, then there exists at least one interval of length $h$, where every token within the interval has a hash value smaller than $H(u_o)$. Thus, we first have the basic probability $\phi^{h-1}$, which represents the probability of finding a watermark window where all tokens' hash values are smaller than $H(u_o)$.

For the case where $X = x$, we need to ensure that this interval can move to the right by at least $x$ positions. Clearly, $i \geq x$ is required to influence the subsequent $x$ tokens. We now consider the suffix, which is the distribution of $\{y_o, \dots, y_{o+h-i-1}\}$. Suppose there are $K$ tokens in the suffix whose hash values are smaller than $H(u_o)$, then these $K = k$ tokens will be affected by the modification of $y_o$ to $u_o$, with probability $P(K = k | I = i) = \phi^{h-1} \cdot (1 - \phi)^{i-k} \phi^k$.

For each affected token, there is a probability of $1 - \gamma$ that it belongs to the red list $R$ (i.e., the watermark is removed), and a probability of $\gamma$ that it remains in the "green" set (i.e., the watermark is not removed). Given $K = k$, the value of $X$ follows a binomial distribution $\text{Binomial}(k, 1 - \gamma)$. The probability of removing $x$ watermarks from the $k$ affected tokens is given by:
\begin{equation}
    P(X = x | K = k) = \binom{k}{x} (1 - \gamma)^x \gamma^{k - x}
\end{equation}
Given $I = i$, for all possible values of $k \in \{x, x+1, \dots, i\}$, we define $P(X = x | I = i)$ as the probability of removing $x$ watermarks, which, by the law of total probability, is:
\begin{equation}
P(X = x | I = i) = \sum_{k = x}^{i} P(X = x | I = i, K = k) = \sum_{k = x}^{i} P(X = x | K = k) P(K = k | I = i)
\end{equation}
Substituting the expressions for $P(X = x | K = k)$ and $P(K = k | I = i)$:
\begin{equation}
P(X = x | I = i) = \phi^{h-1} \times \sum_{k = x}^{i} (1 - \phi)^{i - k} \phi^k \binom{k}{x} (1 - \gamma)^x \gamma^{k - x}
\end{equation}
Similarly, by summing over all possible values of $i \in \{x, x+1, \dots, h\}$, we obtain the final expression for $P(X = x)$ using the law of total probability:
\begin{align}
P(X = x) &= \sum_{i = x}^{h} P(X = x | I = i) \\
 &= \sum_{i = x}^{h} \phi^{h - 1} \times \sum_{k = x}^{i} (1 - \phi)^{i - k} \phi^k \binom{k}{x} (1 - \gamma)^x \gamma^{k - x}\\
 &= \sum_{i = x}^{h} \sum_{k = x}^{i} (1 - \phi)^{i - k} \phi^{k + h - 1} \binom{k}{x} (1 - \gamma)^x \gamma^{k - x}
\end{align}
We further generalize the hypothesis. If we treat $\Phi = \phi$ as a variable, we need to consider all possible values for the inserted token. For simplicity, let $p = \frac{v}{|\mathcal{V}|}$, where $v$ is the index of the token in the vocabulary $\mathcal{V}$:
\begin{align}
\hat{P} &= \sum_{\phi} P(X = x | \Phi = \phi) = \sum_{v = 1}^{|\mathcal{V}|} \sum_{i = x}^{h} \sum_{k = x}^{i} (1 - p)^{i - k} p^{k + h - 1} \binom{k}{x} (1 - \gamma)^x \gamma^{k - x}
\end{align}


\begin{proposition}
For our watermark scheme with a window size of $h$ and hash space $d$, the probability of having exactly $X$ watermarks removed is given by:
\begin{equation}
P(X=x)=\sum_{i=x}^h\sum_{k=x}^i(\frac{1}{d})^{i-k}(1-\frac{1}{d})^{h+k-1}\binom{k}{x}(\frac{1-\gamma}{d})^x(1-\frac{1-\gamma}{d})^{k-x}
\end{equation}
\end{proposition}
\textbf{Proof:} We define the following events:
\begin{itemize}
    \item \textbf{Event A}: The watermark window does not include any texture key to protect the watermark text. 
    \item \textbf{Event B}: The watermark window contains only one texture key to protect the watermark text.
    \item \textbf{Event C}: More than two tokens from the watermark window are involved in the hash space that protects the watermark text. In this case, modifying a single token will not remove the watermark.
\end{itemize}

Events A and C only contribute to $P(X = 0)$. Therefore, when considering perturbations for $X > 0$, we only need to focus on Event B. We define \( I = i \) as the number of consecutive tokens, starting from \( y_{o-1} \), that satisfy the condition \( H(y) \neq H(u_o) \), where \( u_o \) represents the position of the watermark. The sequence \( \{ y_{o-i}, \dots, u_o, \dots, y_{o+h-i-1} \} \) represents the first watermark window, where \( i \) is the maximum sliding distance for this window. The basic probability of a watermark window where no token's hash value equals \( H(u_o) \) is given by $\left( 1 - \frac{1}{d} \right)^{h-1}$,
where \( m \) is the size of the hash space. For \( X = x \), there must exist a segment within the watermark window that can slide to the right by at least \( x \) positions. This implies that \( i \geq x \) for the watermark to affect subsequent tokens. Next, we consider the suffix \( \{ y_o, \dots, y_{o+h-i-1} \} \) and the probability distribution \( P(K=k|I=i) \), where \( K \) denotes the number of tokens in the suffix that do not match \( H(u_o) \). This probability follows:

\begin{equation}
P(K=k|I=i) = \left( 1 - \frac{1}{d} \right)^{h-1} \cdot \left( \frac{1}{d} \right)^{i-k} \cdot \left( 1 - \frac{1}{d} \right)^k
\end{equation}

Let \( Z \) represent the number of tokens that are affected by $u_o$. Given \( Z=z \), the probability \( P(Z=z|K=k) \) follows a binomial distribution:
\begin{equation}
P(Z=z|K=k) = \binom{k}{z} \left( \frac{1}{d} \right)^z \left( 1 - \frac{1}{d} \right)^{k-z}
\end{equation}
For each affected token, there is a probability of \( 1-\gamma \) of being in the red set (i.e., the watermark is erased) and \( \gamma \) for being in the green set (i.e., the watermark remains). Given \( Z=z \), the value \( X \) follows a binomial distribution \( \text{Binomial}(z, 1-\gamma) \), and the probability of erasing \( x \) watermarks is:
\begin{equation}
P(X=x|Z=z) = \binom{z}{x} (1-\gamma)^{x} \gamma^{z-x}
\end{equation}
We now compute the overall probability:
\begin{align}
P(X=x|K=k) &= \sum_{z=x}^k P(X=x|Z=z) P(Z=z|K=k) \\
&=\sum_{z=x}^k \binom{k}{z} \left( \frac{1}{d} \right)^z \left( 1 - \frac{1}{d} \right)^{k-z} \binom{z}{x} (1-\gamma)^{x} \gamma^{z-x}
\end{align}
Using the law of total probability, we can calculate \( P(X=x|I=i) \) by summing over \( k \):
\begin{align}
P(X=x|I=i) &= \sum_{k=x}^{i} P(X=x|K=k) P(K=k|I=i)
\end{align}
Finally, by iterating over all possible values of \( i \), we obtain the total probability for \( X=x \):
\begin{align}
P(X=x) &= \sum_{i=x}^{h} P(X=x|I=i)\\
&=\sum_{i=x}^h\sum_{k=x}^i(\frac{1}{d})^{i-k}(1-\frac{1}{d})^{h+k-1}\sum_{z=x}^k\binom{k}{z}(\frac{1}{d})^z(1-\frac{1}{d})^{k-z} \binom{z}{x} (1-\gamma)^{x} \gamma^{z-x} \\
&=\sum_{i=x}^h\sum_{k=x}^i(\frac{1}{d})^{i-k}(1-\frac{1}{d})^{h+k-1}\binom{k}{x}\sum_{z=x}^k\binom{k-x}{z-x}(\frac{1}{d})^z(1-\frac{1}{d})^{k-z} (1-\gamma)^{x} \gamma^{z-x}
\end{align}
by adjusting the summation variable, setting \( t = z - x \), which implies \( z = t + x \). Consequently, the range of \( t \) is \( 0 \leq t \leq k - x \). Substituting this into the sum, we obtain:
\begin{equation}
P(X=x) =\sum_{i=x}^n\sum_{k=x}^i(\frac{1}{d})^{i-k}(1-\frac{1}{d})^{h+k-1} \left( \frac{1}{d} \right)^x \sum_{t=0}^{k - x} \binom{k - x}{t} \left( \frac{1}{d} \right)^t \left( 1 - \frac{1}{d} \right)^{(k - x) - t} \gamma^t
\end{equation}
This equation can be simplified by applying the binomial expansion for \( \sum_{t=0}^{n} \binom{n}{t} a^t b^{n - t} = (a + b)^n \), leading to:
\begin{equation}
P(X=x) = \sum_{i=x}^{h} \sum_{k=x}^{i} \left( \frac{1}{d} \right)^{i-k} \left( 1 - \frac{1}{d} \right)^{h+k-1} \binom{k}{x} \left( \frac{1-\gamma}{d} \right)^x \left( 1 - \frac{1-\gamma}{d} \right)^{k-x}
\end{equation}
Alternatively, a more efficient derivation can be achieved by directly applying Proposition~\ref{prop:kgw_min}. In this approach, let \( \phi = 1 - \frac{1}{d} \), and define \( \gamma = \frac{1}{d} (1 - r) \) as the probability of successfully eliminating the watermark. The failure probability is then \( 1 - \frac{1 - r}{d} \), from which this can be directly derived.

\begin{proposition}
Let $\hat{E}(X)$ denote the expected perturbation under the KGW-Min strategy in the presence of adversarial tampering, and let $E(X)$ denote the expected perturbation under our proposed method. There exists a threshold pair $(\gamma_c, d_c)$ such that for $d > d_c$ and $\gamma \geq \gamma_c = \frac{1}{d+1}$, then the inequality $\hat{E}(X) > E(X)$ holds.
\end{proposition}
\textbf{Proof:} 
We use $\hat{P}$ and $P$ to denote the probability that the subsequent watermark tokens are perturbed under the KGW-MIN method and our proposed method, respectively, given that an adversarial manipulation occurs. We next aim to establish the existence of threshold values $r_c$ and $m_c$, such that for all $x > 0$, when $\gamma \geq \gamma_c = \frac{1}{d + 1}$ and $d > d_c$, the inequality $\hat{P}(X = x) \geq P(X = x)$ holds. Formally,
\begin{equation}
\begin{aligned}
\hat{P}(X = x) &= \sum_{v=1}^{|\mathcal{V}|} \sum_{i = x}^{h} \sum_{k = x}^{i} (1 - p)^{i - k} p^{k + h - 1} \binom{k}{x} (1 - \gamma)^x \gamma^{k - x} \\
&> \sum_{i = x}^{n} \sum_{k = x}^{i} \left(\frac{1}{d}\right)^{i - k} \left(1 - \frac{1}{d}\right)^{h + k - 1} \binom{k}{x} \left(\frac{1 - \gamma}{d}\right)^x \left(1 - \frac{1 - \gamma}{d}\right)^{k - x}\\
&=P(X = x)
\end{aligned}
\end{equation}

To prove the above inequality holds universally, we first show that when $\gamma \geq \frac{1}{d + 1}$, the inequality:
\begin{equation}
\left(\frac{1 - \gamma}{d}\right)^x \left(1 - \frac{1 - \gamma}{d}\right)^{k - x} < (1 - \gamma)^x \gamma^{k - x}
\end{equation}
holds for all $0 \leq x \leq k \leq h$. This inequality is equivalent to showing that $\left(\frac{1}{d}\right)^x \left(1 - \frac{1 - \gamma}{d}\right)^{k - x} < \gamma^{k - x}$. Taking the logarithm of both sides and subtracting yields the difference function, formally:
\begin{equation}
f(k) = -x \ln d + (k - x) \left[ \ln\left(1 - \frac{1 - \gamma}{d}\right) - \ln \gamma \right]
\end{equation}
Since $d > 1$ and $0 < \gamma < 1$, the term $\ln\left(1 - \frac{1 - \gamma}{d}\right) - \ln \gamma$ is always non-negative. Therefore, $f(k)$ increases with $k$, and we have $f(k) < f(h) < -x \ln d + h \left[ \ln\left(1 - \frac{1 - \gamma}{d}\right) - \ln \gamma \right]$. To ensure that $f(k) < 0$ holds for all $k$, it suffices to impose:
\begin{equation}
-x \ln d + h \left[ \ln\left(1 - \frac{1 - \gamma}{d}\right) - \ln \gamma \right] \leq 0,
\end{equation}
which yields $\gamma \geq \frac{d - 1}{d^{x/h + 1} - 1}$. To guarantee this inequality holds for any $x$, a sufficient condition is $\gamma \geq \frac{1}{1 + m} \geq \frac{d - 1}{1 + d^{x/h + 1}}$. Hence, if $\gamma \geq \frac{1}{d + 1}$, we have $f(k) < -x \ln d+ h \left[ \ln\left(1 - \frac{1 - \gamma}{d}\right) - \ln \gamma \right] < 0$ always holds, completing the proof of the inequality. As a direct consequence, the following inequality between the probabilities holds:
\begin{equation}
\begin{aligned}
P(X = x) &= \sum_{i=x}^n\sum_{k=x}^i(\frac{1}{d})^{i-k}(1-\frac{1}{d})^{h+k-1}\binom{k}{x}(\frac{1-\gamma}{d})^x(1-\frac{1-\gamma}{d})^{k-x} \\
&< \sum_{i=x}^n\sum_{k=x}^i(\frac{1}{d})^{i-k}(1-\frac{1}{d})^{h+k-1}\binom{k}{x}(1-\gamma)^{x} \gamma^{k-x}
\end{aligned}
\end{equation}

Next, we aim to show that there exists a real-valued threshold $m_c$ such that for all $m > m_c$, the following inequality holds:
\begin{equation}
\sum_{i = x}^h \sum_{k = x}^i \left(\frac{1}{d}\right)^{i - k} \left(1 - \frac{1}{d}\right)^{h + k - 1}
< \sum_{v = 1}^{|\mathcal{V}|} \sum_{i = x}^h \sum_{k = x}^i (1 - p)^{i - k} p^{k + h - 1}
\end{equation}
Let us denote the left-hand side as $S_{\text{left}} = \sum_{i = x}^h \sum_{k = x}^{i} \left( \frac{1}{d} \right)^{i-k} \left( 1 - \frac{1}{d} \right)^{h+k-1}$. Given that the vocabulary size $|\mathcal{V}|$ is typically a large integer, the right-hand side can be approximated by an integral, i.e.,
\begin{equation}
S_{\text{right}} = \sum_{v=1}^{|\mathcal{V}|} \sum_{i = x}^h \sum_{k = x}^i (1 - p)^{i - k} p^{k + h - 1} \approx \int_0\textsuperscript{1} \sum_{i = x}^h \sum_{k = x}^i (1 - p)^{i - k} p^{k + h - 1} \, dp.
\end{equation}

By applying the identity of the Beta function with substitutions $a = i - k$ and $b = k + h - 1$, we obtain:
\begin{equation}
\int_0\textsuperscript{1} (1 - p)^a p^b \, dp = B(a + 1, b + 1) = \frac{a! \, b!}{(a + b + 1)!}
= \frac{(i - k)! \, (k + h - 1)!}{(i + h)!}
\end{equation}
If each term on the left-hand side satisfies the inequality$\left( \frac{1}{d} \right)^{i - k} \left( 1 - \frac{1}{d} \right)^{h + k - 1} < \frac{(i - k)! (k + h - 1)!}{(i + h)!}$, then it directly follows that $S_{\text{left}} < S_{\text{right}}$. Taking the natural logarithm of both sides, we obtain:
\begin{equation}
\left( \frac{1}{d} \right)^{i - k} \left( 1 - \frac{1}{d} \right)^{h + k - 1} < \frac{(i - k)! \, (k + h - 1)!}{(i + h)!}
\end{equation}
Taking the natural logarithm of both sides, the inequality becomes:
\begin{equation}
(i - k) \ln \left( \frac{1}{d} \right) + (h + k - 1) \ln \left( 1 - \frac{1}{d} \right)
< \ln \left( \frac{(i - k)! \, (k + h - 1)!}{(i + h)!} \right)
\end{equation}
Define the function:
\begin{equation}
f(m) = (i - k) \ln \left( \frac{1}{d} \right) + (h + k - 1) \ln \left( 1 - \frac{1}{d} \right)
= - (i - k) \ln m + (h + k - 1) \ln \left(1 - \frac{1}{d} \right)
\end{equation}

To further bound the function $f(m)$, we utilize the inequality $\ln(1 - x) < -x - \frac{x^2}{2}$ for $0 < x < 1$, and note that since $m > 1$, it follows that:
\begin{equation}
\ln\left(1 - \frac{1}{d} \right) \leq -\frac{1}{d} - \frac{1}{2d^2} \quad \Rightarrow \quad
f(m) < - (i - k)\ln d - \frac{h + k - 1}{d}
\end{equation}
To lower-bound the right-hand side of the inequality involving factorial terms, we apply a simplified version of Stirling’s approximation, namely $\ln(n!) \geq n \ln n - n$, and obtain:
\begin{equation}
\begin{aligned}
\ln \left( \frac{(i - k)! (k + h - 1)!}{(i + h)!} \right)
&= \ln (i-k)! + \ln (k + h - 1)! - \ln (i + h)! \\
&\geq (i - k)\ln(i - k) + (k + h - 1)\ln(k + h - 1) \\
&\quad - (i + h)\ln(i + h)
\end{aligned}
\end{equation}

We now consider a constructive lower bound for $m$ as follows:
\begin{equation}
d \geq \left\lceil \frac{(i + h)^{2h}}{(i - k)(h + k - 1)} \right\rceil 
\quad \Rightarrow \quad 
\ln d \geq \ln \left( \frac{(i + h)^{2h}}{(i - k)(h + k - 1)} \right)
\end{equation}
This yields the upper bound:
\begin{equation}
\begin{aligned}
- (i - k) \ln m 
&\leq - (i - k) \ln \left( \frac{(i + h)^{2h}}{(i - k)(h + k - 1)} \right) \\
&= (i - k)\left[ \ln(i - k) + \ln(h + k - 1) - 2h \ln(i + h) \right]
\end{aligned}
\end{equation}
Therefore,
\begin{equation}
f(m) < - (i - k)\ln m - \frac{h + k - 1}{m}
\leq (i - k)\left[ \ln(i - k) + \ln(h + k - 1) - 2h \ln(i + h) \right] - \frac{h + k - 1}{m}
\end{equation}
Since $(i - k) \ln(h + k - 1) < (k + h - 1)\ln(k + h - 1)$ and $-2h \ln(i + h) < - (i + h) \ln(i + h)$, it follows that:
\begin{equation}
\begin{aligned}
&(i - k)\left[ \ln(i - k) + \ln(h + k - 1) - 2h \ln(i + h) \right] - \frac{h + k - 1}{m} \\
&\quad < (i - k)\ln(i - k) + (k + h - 1)\ln(k + h - 1) - (i + h)\ln(i + h)
\end{aligned}
\end{equation}
which guarantees that the desired inequality holds. As a result, there exists a constant threshold $m_c$ such that for all $m > m_c$, we have:
\begin{equation}
\sum_{i = x}^h \sum_{k = x}^i \left(\frac{1}{d}\right)^{i - k} \left(1 - \frac{1}{d}\right)^{h + k - 1}
< \sum_{v = 1}^{|\mathcal{V}|} \sum_{i = x}^h \sum_{k = x}^i (1 - p)^{i - k} p^{k + h - 1}
\end{equation}
Combining the previously established inequality $(\frac{1 - \gamma}{m})^x \left(1 - \frac{1 - \gamma}{m}\right)^{k - x} < (1 - \gamma)^x \gamma^{k - x}$, we obtain the following relationship:
\begin{equation}
\begin{aligned}
\hat{P}(X = x) &= \sum_{v = 1}^{|\mathcal{V}|} \sum_{i = x}^{h} \sum_{k = x}^{i} (1 - \phi)^{i - k} \phi^{k + h - 1} \binom{k}{x} (1 - \gamma)^x \gamma^{k - x} \\
&> \sum_{i = x}^{n} \sum_{k = x}^{i} \left(\frac{1}{d}\right)^{i - k} \left(1 - \frac{1}{d} \right)^{h + k - 1} \binom{k}{x} \left(\frac{1 - \gamma}{d} \right)^x \left(1 - \frac{1 - \gamma}{d} \right)^{k - x}\\& >\sum_{i=x}^n\sum_{k=x}^i(\frac{1}{d})^{i-k}(1-\frac{1}{d})^{h+k-1}\binom{k}{x}(1-\gamma)^{x} \gamma^{k-x}\\& = P(X = x).
\end{aligned}
\end{equation}
Since the inequality $\hat{P}(X = x) > P(X = x)$ holds for all $x > 0$, it must follow that $\hat{P}(X = 0) < P(X = 0)$ to preserve the total probability mass. Let $\hat{E}$ and $E$ denote the expected values under $\hat{P}$ and $P$, respectively. Then,
\begin{equation}
\hat{E}(X) = \sum_{x = 0}^{h} \hat{P}(X = x) \cdot x = \sum_{x = 1}^{h} \hat{P}(X = x) \cdot x, \quad
E(X) = \sum_{x = 0}^{h} P(X = x) \cdot x = \sum_{x = 1}^{h} P(X = x) \cdot x
\end{equation}
Hence, we conclude:
\begin{equation}
\hat{P}(X = x) > P(X = x) \quad \Rightarrow \quad \hat{E}(X) > E(X)
\end{equation}

\textbf{In conclusion}, we have shown that under the conditions $\gamma \geq \frac{1}{d + 1}$ and $d > d_c$, the KGW-MIN mechanism yields a strictly greater expected number of perturbed tokens than the baseline, i.e., $\hat{E}(X) > E(X)$.

\section{Visualization}

To better illustrate the effectiveness of our watermarking method, we provide qualitative visualizations based on representative examples from the C4-RealNewsLike dataset. As shown in Table~\ref{tab:demo-examples2} and Table~\ref{tab:demo-examples}, we present side-by-side comparisons of real completions, unwatermarked generations (NW), and watermarked generations (W) under the same prompt. In the visualization presented in Table~\ref{tab:examples3}, we comparatively analyze the output of unwatermarked text versus two forms of adversarial attacks on the MMW-FakeNews dataset. 
For scrubbing, a higher residual $z$-score post-attack indicates stronger watermark retention. For spoofing, a lower $z$-score indicates higher robustness to imitation. The p-value quantifies the probability that a text with the observed watermark strength (or stronger) could have been generated by chance under the null (unwatermarked) distribution. Empirically, our method demonstrates superior resilience in both cases. Under scrubbing attacks, our watermark yields a substantially higher residual $z$-score (7.38) compared to KGW’s (4.82), despite both being attacked under the same conditions. This suggests that our method better preserves the watermark signal post-perturbation. Additionally, the corresponding p-values further confirm the statistical significance of the retained signal. In the spoofing setting, the attacker on KGW achieves a high $z$-score of 11.09, indicating a successful watermark fabrication. In contrast, spoofed text targeting our watermark only achieves a $z$-score of 2.29, substantially closer to the null distribution. This implies our watermark is significantly more resistant to imitation and cannot be easily spoofed into a falsely high-confidence detection.

\begin{table*}[t]
\tiny
\resizebox{\textwidth}{!}{
\begin{tabular}{p{2.2cm}|p{2.2cm}|p{2.2cm}|p{2.2cm}|c|c}
\toprule
prompt & real completion &  no watermark (NW) &  watermarked (W) & (NW)$z$ & (W) $z$   \\
\midrule
 ...of Wrigley Field's right field video board.\textbackslash nOn Monday morning's "Mully and Hanley" show on WSCR-AM 670, Cubs president of business operations Crane Kenney said the ball is back to where it originally landed after it was taken down after &  the 2015 season.\textbackslash nKenney said the ball was taken down in part to prevent anyone from getting the idea of climbing up the board and stealing it.\textbackslash nMeanwhile, the metal detectors are all in place and finishing touches were being put  [...continues] &   Game 4 last weekend in St. Louis. The Cardinals won that game by a score of 9-3 to advance to the NLCS against the Dodgers.\textbackslash nSchwarber homered off Adam Wainwright with one out in the bottom of the first inning at [...continues] &  Game 1 last Tuesday night.\textbackslash nKenney didn't know who took possession of the baseball or how long before its return that decision had occurred. But he did say that he would like to see more balls come off the[...continues] &  0.27 &    6.61     \\\midrule
 ... on the planet this week at the NFL Scouting Combine. \textbackslash nIt’s an experience that might humble some. But for Butler, it has only enhanced his confidence.\textbackslash nAs it stands, 22-year-old Butler is not regarded as the &    best wide receiver in this year’s NFL Draft. He’s projected by some experts to go as late as the third round. But when wide receivers were measured Thursday, Butler gained some attention: He led all receivers in height (6-foot-5 3/8) [...continues] &   top receiver in this year’s draft class. He doesn’t have the speed or quickness to match up with most cornerbacks and he isn’t big enough to dominate against safeties.\textbackslash nButler knows all about those shortcomings.[...continues] & top receiver in this class and he isn’t expected to go until Day Two or three. The Iowa State standout has drawn comparisons with Michael Crabtree because both were tall and lanky pass catchers who didn’t blow up the combine but   [...continues] &  0.54 &      11.21    \\\midrule
 ...you’ll be able to walk around a person inside a virtual environment to see their entire being. The Wellington, New Zealand-based startup said its VR platform will allow you to view fully volumetric 3D video.\textbackslash n &   Previously, VR cameras could record a partial torso of a person, but the 8i Portal goes beyond that. It enables “complete freedom of movement,” and it gives you a sense of “presence,” or the feeling that you are there in the virtual[...continues] &   The company is also launching an SDK that allows developers to integrate 8i into apps on mobile devices or desktop computers. It can work with any headset including Oculus Rift, HTC Vive, Samsung GearVR,[...continues] &   The company raised \$10 million last year from investors such as Peter Thiel and Yuri Milner.\textbackslash nWithin minutes I could step into an office with my friend and look at him full on or through his glasses. We were both wearing  Oculus Rift headsets [...continues] &     0.28 &     14.40       \\\midrule
...The five-member board is expected to support the plan, which was announced amid much hoopla on Nov. 13. The proposed county incentives are part of an agreement in which Amazon would occupy significant office &  space and bring at least 25,000 high-paying jobs to Arlington in coming years.\textbackslash nOpponents hope to postpone the vote until after additional public hearings, where they want representatives of the online retail giant to answer questions directly from anyone in [...continues] &  space at two buildings under construction by JBG Smith Properties Inc., one of Washington’s largest real estate companies.\textbackslash nAmazon has said it will bring as many as 25,000 jobs paying an average salary of more than \$150,000 over time — with some starting out making less[...continues] & space at two properties owned by JBG Smith: PenPlace (near Reagan National Airport) and Potomac Yard (on U.S. Route 1).\textbackslash nJBG has committed more than \$500 million in capital improvements at both sites that include new transportation[...continues] &     0.21 & 8.96       \\
\bottomrule
\end{tabular}
}
\caption{Selected outputs from non-watermarked (NW) and watermarked (W) multinomial sampling using $\gamma=0.25$ and $\delta=2.0$. 
}
\label{tab:demo-examples2}
\vspace{-.2cm}
\end{table*}

\begin{table}[t]
\centering
\caption{\small Visualization of watermarked text samples from the MMW-FakeNews dataset under scrubbing and spoofing attacks. Compared are the KGW watermark and our proposed method. A smaller p-value provides stronger statistical evidence against the null hypothesis. Green and red tokens are color-coded, respectively.}
\resizebox{\textwidth}{!}{
\begin{tabular}{p{3.8cm} | m{13cm}}
\toprule
\centering Prompt & Write a news article about Narendra Modi's visit to Denis Sassou Nguesso in a space exploration symposium. It is very important to never output very short texts, below 300 words... \\
\midrule
\makecell{\textbf{Unwatermarked} \\ p-value: 0.53 \\ $z$-score: -0.08}  & \red{Title: Nar\green{endra} \green{Mod}i's Historic Visit to \green{Den}is S\green{assou} Nguesso \green{in} Space Exploration Symposium\texttt{\textbackslash n\textbackslash n} Introduction: \green{Ind}ia's Prime Minister Narendra \green{Mod}i made a \green{groundbreaking} visit to the \green{President} of \green{the} Republic of Congo, \green{Denis} Sassou Nguesso, during a space exploration symposium \green{held} in \green{the} capital city of \green{Bra}zzaville. This visit marks a significant} [continues...] \\
\midrule
\makecell{\textbf{(a) Scrubbing Attacked} \\ (KGW) \\ p-value: 6.89e-07\\ $z$-score: 4.82~($\downarrow$ 10.47)} &  \red{Title: Modi-S\green{ass}ou \green{N}gu\green{esso Summit Highlight}s Ind\green{o}-Congol\green{ese} \green{Space} Exploration Part\green{ners}hip\texttt{\textbackslash n\textbackslash n} \green{Introduction}:\texttt{\textbackslash \green{n}}  India'\green{s} Prime \green{Minister Nar}endra Modi and \green{the} President of \green{Congo Den}is \green{Sass}ou Nguesso recently met at \green{a} space exploration sympos\green{ium in} New \green{Del}hi, \green{signal}ing \green{a} new era in Indo-Congolese collaboration \green{in} the \green{field}. This historic} [continues...]  \\
\midrule
\makecell{\textbf{(b) Scrubbing Attacked} \\ (Ours) \\ p-value: 8.59e-14 \\ $z$-score: 7.38~($\downarrow$ 6.72)} &  \red{On Friday, Indian Prime \green{Minister} Nar\green{endra Mod}i met \green{with} \green{President Denis} S\green{assou} Nguesso \green{during a} symposium \green{focused} on \green{space exploration}\green{. Mod}i was in \green{France} for several official \green{engag}ements, \green{including} a bilateral discussion \green{with French} President Emmanuel Mac\green{ron}. The event, titled **"Reaching \green{for the} Stars: A New \green{Frontier} \green{in} Space Ex\green{pl}oration,"} [continues...]  \\
\midrule
\makecell{\textbf{(c) Spoofing Attacked} \\ (KGW) \\ p-value: 4.72e-39 \\ $z$-score: 11.09} &  \red{Title: \green{M}elting Bound\green{aries:} \green{Nar}endra \green{Mod}i's \green{Historic} Visit to Den\green{is} \green{Sass}ou \green{N}gu\green{esso}'s \green{Space} \green{Expl}oration \green{Sym}pos\green{ium} \texttt{\green{\textbackslash n}\green{\textbackslash n}} \green{Introduction:} On \green{a} sunny \green{day} in the \green{month of March}, \green{the} world \green{witnessed a} \green{significant} historical \green{event} when \green{Prime} Minister \green{Nar}endra \green{Mod}i of India \green{paid} \green{a} \green{visit} \green{to} the \green{President of} Republic \green{of} \green{the Congo}, Den\green{is} \green{Sassou} Nguesso, at \green{the} latter\green{'s}}[continues...] \\
\midrule
\makecell{\textbf{(d) Spoofing Attacked} \\ (Ours) \\ p-value: 0.01 \\ $z$-score: 2.29} & \red{Title: \green{Mod}i's Historic Visit to \green{Den}is S\green{assou} Nguesso \green{in} Space Ex\green{pl}oration Symposium \texttt{\textbackslash n\textbackslash n} Introduction: \green{As} \green{the} world continues to \green{evolve} and \green{push} the \green{boundaries} of technology, \green{the} race to explore and conquer space has \green{gained} significant momentum. \green{In} this context, \green{the} recent visit of \green{Indian} Prime Minister Narendra Modi to \green{Den}is Sas\green{sou} Nguessa, the} [continues...]
 \\
\bottomrule
\end{tabular}
}
\label{tab:examples3}
\end{table}

\section{Impletement Details}
\label{sec:Exp_details}
\subsection{Detailed Experimental Setup}

\paragraph{Dataset and prompts.} 
Following previous work \citep{KWG1, kuditipudi2023robust, steal}, we utilize the RealNewsLike subset of the C4 dataset \citep{c4}, WikiText-103 \citep{wikitext}, and other datasets that support open-ended text generation to construct various prompts. For the human evaluation study, we select the Long-Form Question Answering (LFQA) dataset \citep{dipper} to assess the robustness of the watermarking method across real-world question-answering scenarios. During generation, we randomly sample text segments from the dataset and truncate a fixed-length portion from the end, using the remainder as the prompt for model completion. Following the setup in~\citet{KWG1,KWG2}, all experimental results presented in the figures and tables are based on the same 500 positive samples and 500 negative samples. For the spoofing attack, the attacker uses query prompts derived from the C4-RealNewsLike dataset to obtain 30,000 responses from the victim. Each response contains fewer than 800 tokens.

\paragraph{Hyper-parameters.} Unless otherwise specified, the watermarking schemes in our experiments adopt the hyperparameter settings commonly used in prior work~\citep{KWG1,KWG2,steal}, (\(\gamma = 0.25\), \(\delta = 5\)) , and a maximum generation length of 150 new tokens. In the spoofing attack experiments, all the watermark detectors are calibrated on the C4-RealNewsLike dataset using 2,000 watermarked and non-watermarked texts. Model outputs containing fewer than 150 tokens are discarded. For the spoofing learning, the attacker generates original queries prompt using the C4-RealNewsLike subset, obtaining no fewer than n=30,000 responses, each with a maximum token length of 800. The spoofing model is configured with a spoofer strength of 8.25, and a weighted loss objective defined by $w_{\text{abcd}} = 2.0$, $w_{\text{partials}} = 1.0$,  $w_{\text{empty}} = 0.5$. All experimental results presented in the figures and table are based on more than 500 positive samples and over 500 negative samples. Unless otherwise specified, \model~in this paper uses the hyperparameters d=6 and h=6. All experiments are conducted on Nvidia A40 GPUs.

\subsection{Configuration LLM Scrubbing Attack}

\begin{tcolorbox}[title = {Prompt for Scrubbing Attack Specifications}]
\{\textcolor{purple}{Prompt}\} As an expert copy-editor, please rewrite the following text in your own voice while ensuring that the final output contains the same information as the original text and has roughly the same length. Please paraphrase all sentences and do not omit any crucial details. Additionally, please take care to provide any relevant information about public figures, organizations, or other entities mentioned in the text to avoid any potential misunderstandings or biases. \{\textcolor{blue}{Watermark Paragraph}\}......\\
\end{tcolorbox}

\begin{table*}[t]
\tiny
\resizebox{\textwidth}{!}{
\begin{tabular}{p{2.2cm}|p{2.2cm}|p{2.2cm}|p{2.2cm}|c|c}
\toprule
prompt & real completion &  no watermark (NW) &  watermarked (W) & (NW)$z$ & (W) $z$   \\
\midrule
 ...Shania Twain expected to break the charts with new album NOW! \textbackslash nEven after a 15-year hiatus, she’s still the one! Shania Twain is on pace to top the charts with her new album NOW. The 16-song LP was released on Sept. 29 &  the 2015 season.\textbackslash nTimmins native hopped on Twitter on Wednesday to announce that her album is already platinum in Canada!\textbackslash nShortly after the release of the lead single “Life’s About to Get Good,” it hit number one[...continues] &   album is expected to sell 100,000 copies in its first week, according to Hits Daily Double.\textbackslash nShania, 49, has been teasing the release of NOW for months. In June, she released the album’s first single, “Life’s About to Get Good,”[...continues] &  album is currently at number two and has earned more than\$111K. \textbackslash nThe pop/country powerhouse has released five number one records and has earned five Grammys.\textbackslash nShania has released three albums in English and two in French.[...continues] &  0.38 &    21.17     \\\midrule
 ... Covering more than 30 auto manufacturers worldwide and providing more than 10 years of historical data, the 32 new factors are designed to assist in the prediction of stock returns using analytics derived from company &    specific datapoints on sales, production and market share. They include multiple financial factors for revenue and sales as well as novel factors covering production of electric vehicles and plant utilization.[...continues] &  specific news and social media. \textbackslash nThe new signals are available through IHS Markit’s Alternative Data Hub, which provides access to more than 100 alternative data signals across 10 sectors. “The automotive secto [...continues] & specific news and social media content. The new factor suite offers clients additional insight and context for understanding market movements in this sector.The new suite has two two components. The first [...continues] &  0.49 &   22.91    \\\midrule

 ... When I was in my 20s and early 30s, my whole life was focused on work. I didn’t take vacations or weekends off. I was always the first in the office and the last to leave. These days, I’m better &   at balancing the work that I love to do with my foundation and taking time off to spend with family and friends. My parents first taught me bridge, but I really started to enjoy it after[...continues] &   at balancing my work and personal life, but I still have a lot to learn. I’ve been fortunate to meet some amazing people who have helped me along the way. One of them is Roger Federer. I’ve known Roger for [...continues] &   at balancing my life. But I know that I could still stand to disconnect more.
So last week I took two days off. I didn’t touch my Microsoft account or my iPhone. No e-mail. No Twitter. No social media. [...continues] &   -0.91 &     21.47      \\\midrule

... high definition screen projects an intense college basketball game. Massage therapists rub the nervous tensions of men and women away. Scissors skillfully cut men's hair. Two chandeliers adorn the main room, complimented by brick &  walls and a glass bar that doubles as a retail counter. Sean Heywood, right, and Kumi Walker own MR., a barbershop and wine bar in San Francisco, California. This is not your typical barbershop. And that has always been the vision of owners[...continues] &  walls and wooden floors. This isn't your typical barbershop. It's the Barbershop Museum in Tulsa, Oklahoma, and it's the only one of its kind in the United States.
The museum is a tribute to the African-American barbershop, a place[...continues] & walls and rich colors. The only thing reminiscent of an old photo in "Black History: The Legend and The Legendaries" at The Legendaries Salon and Bistro in downtown Chicago is one woman. Dressed in an all green suit with gold accents[...continues] &     -0.53 & 22.61       \\
\bottomrule
\end{tabular}
}
\caption{Selected outputs from non-watermarked (NW) and watermarked (W) multinomial sampling using $\gamma=0.25$ and $\delta=5.0$. 
}
\label{tab:demo-examples}
\vspace{-.2cm}
\end{table*}

\subsection{Full Datasets and Baseline}
\label{appendix:datasets}
\paragraph{Scrubbing prompt dataset}
\begin{enumerate}
\item \textbf{C4.} The Colossal Clean Crawled Corpus (C4) is a large-scale, English-language dataset constructed by applying extensive cleaning and filtering to the Common Crawl web scrape. Introduced by Raffel et al.~\citep{c4} in the context of the T5 framework, C4 contains hundreds of gigabytes of naturalistic web text after removing boilerplate, navigation, and low-quality content. It serves as a comprehensive corpus for pretraining large-scale language models due to its linguistic diversity and domain variability. C4-RealNewsLike is a filtered subset of the original C4 corpus designed to more closely resemble high-quality journalistic writing. This variant was introduced to improve alignment with tasks requiring formal, factually grounded language, such as long-form question answering and summarization. 

\item \textbf{WikiText.} WikiText-103 is a high-quality, curated corpus of English Wikipedia articles introduced by Merity et al.~\citep{wikitext}, specifically constructed for the purpose of training and evaluating autoregressive language models. Unlike raw Wikipedia dumps, which often contain noisy or fragmented content, WikiText-103 retains the full article structure, including paragraph breaks and sequential sentence order, thereby enabling more realistic modeling of long-range dependencies in natural language. The dataset comprises approximately 103 million tokens and has become a widely adopted benchmark for assessing the linguistic coherence, contextual reasoning, and generative fluency of modern language models.

\item \textbf{LFQA.} The LFQA dataset is derived from the ELI5 (Explain Like I’m Five) corpus~\citep{eli5_lfqa} and was introduced to support long-form question answering tasks. It comprises open-domain questions sourced from Reddit, paired with multi-sentence, explanatory answers written in natural language. The dataset emphasizes reasoning, coherence, and knowledge synthesis, making it suitable for training and evaluating models intended for complex, multi-hop generative QA.

\end{enumerate}

\paragraph{Spoofing prompt dataset}
\begin{enumerate}
\item \textbf{Dolly-CW.} We utilize the Dolly dataset~\citep{DollyV2}, which comprises a collection of instruction–response pairs designed to facilitate instruction tuning of large language models. The dataset includes a broad range of tasks spanning classification, generation, information retrieval, and creative writing. Notably, the data was generated using open-source models and later filtered for quality, offering a diverse yet structured corpus suitable for aligning models with human-style instruction following. Following the setup proposed by \citep{steal}, we construct a set of diverse prompts by selecting a representative subset of the Dolly corpus. This subset emphasizes multi-domain scenarios such as question answering, summarization, reasoning, and creative text generation. The prompts are designed to reflect realistic and varied user intents, enabling robust evaluation.

\item \textbf{MMW.}: The MarkMyWords (MMW) dataset~\citep{piet2023mark} is a benchmark designed to evaluate the effectiveness of watermarking schemes for large language model outputs across natural language generation tasks. It comprises three core tasks—BookReports, FakeNews, and StoryGeneration, which are designed to reflect realistic misuse scenarios and to support the evaluation of watermark quality, detectability, and tamper resistance.

The BookReports component includes 500 prompts instructing the model to generate analytical or descriptive reports on well-known books. Each generation consists of long-form text up to 1024 tokens, simulating academic scenarios where LLMs might be misused for student assignments. The outputs provide a structured, content-rich context for evaluating watermark detection and robustness in educational settings. 

The FakeNews component comprises 500 prompts directing the model to fabricate news articles about political figures and fictional events. This task emulates potential misuse of LLMs in disinformation or propaganda campaigns. The generated outputs exhibit journalistic style, incorporate named entities, and follow coherent narrative structures, providing a high-risk setting for testing watermark resilience under adversarial conditions.

\end{enumerate}

\section{Limitation}
\label{section_limitation}
Although our empirical analysis demonstrates that the \model~is effective across a wide range of scenarios and against various watermarking attacks, it is important to acknowledge its limitations. Previous work suggests a potential spoofing attack, in which an attacker may generate a forged text with a high z-score by rearranging or inserting content into an existing watermarked text. While such spoofing attacks do not enable the automatic generation of topic-specific content, current watermarking methods lack robust defenses against these forms of spoofing. The core limitation lies in the fact that the statistical watermarking technique, represented by KGW, is applied at the paragraph level, offering no protection against localized changes. As a result, it cannot detect or prevent modifications to specific portions of the text. This inherent weakness is present in all current watermarking approaches. Addressing this challenge requires the exploration of alternatives to z-test statistical methods, such as integrating semantic or syntactic watermarking techniques, utilizing sentence-level perturbations, or embedding watermark signals deeper within the model.


\section{Ethical Impacts}
The rapid development of large language models has significantly enhanced text generation capabilities, enabling the production of highly human-like content. Watermarking techniques have emerged as an effective approach to mitigate the risks associated with the misuse of LLMs, such as spreading misinformation, plagiarism, and copyright infringement.

Most existing watermarking methods involve a trade-off between scrubbing robustness and spoofing resistance. Recent studies show that if the green and red token sets used in watermarking are compromised via extraction attacks, both robustness and security can degrade substantially. Adversaries can break existing watermarking schemes by statistically analyzing token distributions within the watermarking window in preprint outputs. Such attacks allow malicious users to either erase watermarks from benign content or inject watermarked, toxic content to falsely implicate a target LLM. At a fundamental level, these attacks undermine the ability of watermarking to reliably distinguish machine-generated content from human-authored text, threatening the societal trust. 

Our goal is to develop a watermarking algorithm that mitigates the risk of spoofing under statistical attacks while maintaining strong scrubbing robustness, thereby offering a more reliable and trustworthy watermarking solution for LLMs.

\section{Broader Impacts}
\label{section_impacts}
\paragraph{Positive Societal Impacts.}

The proposed \model~ watermarking framework enhances the resilience of large language models against two critical forms of adversarial manipulation: scrubbing and spoofing. By improving watermark detection without compromising generation quality, SEEK provides a promising tool for responsible deployment of generative models, particularly in contexts where provenance, authenticity, and misuse prevention are essential—such as combating misinformation, enforcing intellectual property rights, and supporting academic integrity. Furthermore, the parameter-efficient and architecture-agnostic nature of the method may help democratize watermarking technology, making it more accessible to practitioners in low-resource settings.

\paragraph{Negative Societal Impacts.}
While watermarking technology is designed to preserve text generation quality, watermarking inherently perturbs token selection probabilities to embed identifiable signals. In high-stakes domains such as medical or legal language modeling, even minor deviations from optimal token choices may introduce semantic ambiguities or factual inaccuracies. As such, watermark-induced perturbations, though imperceptible in general settings, could undermine output fidelity in tasks requiring high precision and domain-specific consistency. This raises concerns about deploying watermarking methods in safety-critical or sensitive applications without extensive domain-specific evaluation.


\newpage

\section*{NeurIPS Paper Checklist}

\begin{enumerate}

\item {\bf Claims}
    \item[] Question: Do the main claims made in the abstract and introduction accurately reflect the paper's contributions and scope?
    \item[] Answer: \answerYes{} 
    \item[] Justification: In the abstract and introduction sections, we articulate the motivation behind \model~, highlight its differences from previous work, and outline the contributions of this paper.
    \item[] Guidelines:
    \begin{itemize}
        \item The answer NA means that the abstract and introduction do not include the claims made in the paper.
        \item The abstract and/or introduction should clearly state the claims made, including the contributions made in the paper and important assumptions and limitations. A No or NA answer to this question will not be perceived well by the reviewers. 
        \item The claims made should match theoretical and experimental results, and reflect how much the results can be expected to generalize to other settings. 
        \item It is fine to include aspirational goals as motivation as long as it is clear that these goals are not attained by the paper. 
    \end{itemize}

\item {\bf Limitations}
    \item[] Question: Does the paper discuss the limitations of the work performed by the authors?
    \item[] Answer: \answerYes{} 
    \item[] Justification: We discuss limitations in the Appendix \ref{section_limitation}.
    \item[] Guidelines:
    \begin{itemize}
        \item The answer NA means that the paper has no limitation while the answer No means that the paper has limitations, but those are not discussed in the paper. 
        \item The authors are encouraged to create a separate "Limitations" section in their paper.
        \item The paper should point out any strong assumptions and how robust the results are to violations of these assumptions (e.g., independence assumptions, noiseless settings, model well-specification, asymptotic approximations only holding locally). The authors should reflect on how these assumptions might be violated in practice and what the implications would be.
        \item The authors should reflect on the scope of the claims made, e.g., if the approach was only tested on a few datasets or with a few runs. In general, empirical results often depend on implicit assumptions, which should be articulated.
        \item The authors should reflect on the factors that influence the performance of the approach. For example, a facial recognition algorithm may perform poorly when image resolution is low or images are taken in low lighting. Or a speech-to-text system might not be used reliably to provide closed captions for online lectures because it fails to handle technical jargon.
        \item The authors should discuss the computational efficiency of the proposed algorithms and how they scale with dataset size.
        \item If applicable, the authors should discuss possible limitations of their approach to address problems of privacy and fairness.
        \item While the authors might fear that complete honesty about limitations might be used by reviewers as grounds for rejection, a worse outcome might be that reviewers discover limitations that aren't acknowledged in the paper. The authors should use their best judgment and recognize that individual actions in favor of transparency play an important role in developing norms that preserve the integrity of the community. Reviewers will be specifically instructed to not penalize honesty concerning limitations.
    \end{itemize}
    
\item {\bf Theory Assumptions and Proofs}
    \item[] Question: For each theoretical result, does the paper provide the full set of assumptions and a complete (and correct) proof?
    \item[] Answer: \answerYes{} 
    \item[] Justification: We discuss all the theory assumptions and proofs in the Appendix \ref{sec:roubutness_theroy} and \ref{sec:method}.
    \item[] Guidelines:
    \begin{itemize}
        \item The answer NA means that the paper does not include theoretical results. 
        \item All the theorems, formulas, and proofs in the paper should be numbered and cross-referenced.
        \item All assumptions should be clearly stated or referenced in the statement of any theorems.
        \item The proofs can either appear in the main paper or the supplemental material, but if they appear in the supplemental material, the authors are encouraged to provide a short proof sketch to provide intuition. 
        \item Inversely, any informal proof provided in the core of the paper should be complemented by formal proofs provided in appendix or supplemental material.
        \item Theorems and Lemmas that the proof relies upon should be properly referenced. 
    \end{itemize}
    
    \item {\bf Experimental Result Reproducibility}
    \item[] Question: Does the paper fully disclose all the information needed to reproduce the main experimental results of the paper to the extent that it affects the main claims and/or conclusions of the paper (regardless of whether the code and data are provided or not)?
    \item[] Answer: \answerYes{} 
    \item[] Justification: We provide a detailed explanation of the parameter usage and will release the source code to ensure reproducibility.
    \item[] Guidelines:
    \begin{itemize}
        \item The answer NA means that the paper does not include experiments.
        \item If the paper includes experiments, a No answer to this question will not be perceived well by the reviewers: Making the paper reproducible is important, regardless of whether the code and data are provided or not.
        \item If the contribution is a dataset and/or model, the authors should describe the steps taken to make their results reproducible or verifiable. 
        \item Depending on the contribution, reproducibility can be accomplished in various ways. For example, if the contribution is a novel architecture, describing the architecture fully might suffice, or if the contribution is a specific model and empirical evaluation, it may be necessary to either make it possible for others to replicate the model with the same dataset, or provide access to the model. In general. releasing code and data is often one good way to accomplish this, but reproducibility can also be provided via detailed instructions for how to replicate the results, access to a hosted model (e.g., in the case of a large language model), releasing of a model checkpoint, or other means that are appropriate to the research performed.
        \item While NeurIPS does not require releasing code, the conference does require all submissions to provide some reasonable avenue for reproducibility, which may depend on the nature of the contribution. For example
        \begin{enumerate}
            \item If the contribution is primarily a new algorithm, the paper should make it clear how to reproduce that algorithm.
            \item If the contribution is primarily a new model architecture, the paper should describe the architecture clearly and fully.
            \item If the contribution is a new model (e.g., a large language model), then there should either be a way to access this model for reproducing the results or a way to reproduce the model (e.g., with an open-source dataset or instructions for how to construct the dataset).
            \item We recognize that reproducibility may be tricky in some cases, in which case authors are welcome to describe the particular way they provide for reproducibility. In the case of closed-source models, it may be that access to the model is limited in some way (e.g., to registered users), but it should be possible for other researchers to have some path to reproducing or verifying the results.
        \end{enumerate}
    \end{itemize}

\item {\bf Open access to data and code}
    \item[] Question: Does the paper provide open access to the data and code, with sufficient instructions to faithfully reproduce the main experimental results, as described in supplemental material?
    \item[] Answer: \answerYes{} 
    \item[] Justification: We provide a detailed explanation of parameter usage and will release the source code.
    \item[] Guidelines:
    \begin{itemize}
        \item The answer NA means that paper does not include experiments requiring code.
        \item Please see the NeurIPS code and data submission guidelines (\url{https://nips.cc/public/guides/CodeSubmissionPolicy}) for more details.
        \item While we encourage the release of code and data, we understand that this might not be possible, so “No” is an acceptable answer. Papers cannot be rejected simply for not including code, unless this is central to the contribution (e.g., for a new open-source benchmark).
        \item The instructions should contain the exact command and environment needed to run to reproduce the results. See the NeurIPS code and data submission guidelines (\url{https://nips.cc/public/guides/CodeSubmissionPolicy}) for more details.
        \item The authors should provide instructions on data access and preparation, including how to access the raw data, preprocessed data, intermediate data, and generated data, etc.
        \item The authors should provide scripts to reproduce all experimental results for the new proposed method and baselines. If only a subset of experiments are reproducible, they should state which ones are omitted from the script and why.
        \item At submission time, to preserve anonymity, the authors should release anonymized versions (if applicable).
        \item Providing as much information as possible in supplemental material (appended to the paper) is recommended, but including URLs to data and code is permitted.
    \end{itemize}

\item {\bf Experimental Setting/Details}
    \item[] Question: Does the paper specify all the training and test details (e.g., data splits, hyperparameters, how they were chosen, type of optimizer, etc.) necessary to understand the results?
    \item[] Answer: \answerYes{} 
    \item[] Justification: The full details are provided in the appendix and supplemental material (code).
    \item[] Guidelines:
    \begin{itemize}
        \item The answer NA means that the paper does not include experiments.
        \item The experimental setting should be presented in the core of the paper to a level of detail that is necessary to appreciate the results and make sense of them.
        \item The full details can be provided either with the code, in appendix, or as supplemental material.
    \end{itemize}

\item {\bf Experiment Statistical Significance}
    \item[] Question: Does the paper report error bars suitably and correctly defined or other appropriate information about the statistical significance of the experiments?
    \item[] Answer:  \answerYes{} 
    \item[] Justification: Experimental results are tested multiple times to ensure stability and reliability.
    \item[] Guidelines:
    \begin{itemize}
        \item The answer NA means that the paper does not include experiments.
        \item The authors should answer "Yes" if the results are accompanied by error bars, confidence intervals, or statistical significance tests, at least for the experiments that support the main claims of the paper.
        \item The factors of variability that the error bars are capturing should be clearly stated (for example, train/test split, initialization, random drawing of some parameter, or overall run with given experimental conditions).
        \item The method for calculating the error bars should be explained (closed form formula, call to a library function, bootstrap, etc.)
        \item The assumptions made should be given (e.g., Normally distributed errors).
        \item It should be clear whether the error bar is the standard deviation or the standard error of the mean.
        \item It is OK to report 1-sigma error bars, but one should state it. The authors should preferably report a 2-sigma error bar than state that they have a 96\% CI, if the hypothesis of Normality of errors is not verified.
        \item For asymmetric distributions, the authors should be careful not to show in tables or figures symmetric error bars that would yield results that are out of range (e.g. negative error rates).
        \item If error bars are reported in tables or plots, The authors should explain in the text how they were calculated and reference the corresponding figures or tables in the text.
    \end{itemize}
    
\item {\bf Experiments Compute Resources}
    \item[] Question: For each experiment, does the paper provide sufficient information on the computer resources (type of compute workers, memory, time of execution) needed to reproduce the experiments?
    \item[] Answer: \answerYes{}
    \item[] Justification: Compute resources are thoroughly described and evaluated in both the experimental setup and the experimental results sections.
    \item[] Guidelines:
    \begin{itemize}
        \item The answer NA means that the paper does not include experiments.
        \item The paper should indicate the type of compute workers CPU or GPU, internal cluster, or cloud provider, including relevant memory and storage.
        \item The paper should provide the amount of compute required for each of the individual experimental runs as well as estimate the total compute. 
        \item The paper should disclose whether the full research project required more compute than the experiments reported in the paper (e.g., preliminary or failed experiments that didn't make it into the paper). 
    \end{itemize}

\item {\bf Code Of Ethics}
    \item[] Question: Does the research conducted in the paper conform, in every respect, with the NeurIPS Code of Ethics \url{https://neurips.cc/public/EthicsGuidelines}?
    \item[] Answer: \answerYes{} 
    \item[] Justification: The research conducted in this paper fully conforms to the NeurIPS Code of Ethics in every respect.
    \item[] Guidelines:
    \begin{itemize}
        \item The answer NA means that the authors have not reviewed the NeurIPS Code of Ethics.
        \item If the authors answer No, they should explain the special circumstances that require a deviation from the Code of Ethics.
        \item The authors should make sure to preserve anonymity (e.g., if there is a special consideration due to laws or regulations in their jurisdiction).
    \end{itemize}

\item {\bf Broader Impacts}
    \item[] Question: Does the paper discuss both potential positive societal impacts and negative societal impacts of the work performed?
    \item[] Answer: \answerYes{}  
    \item[] Justification: We discuss the potential positive societal impacts and negative societal impacts of the work performed in Appendix~\ref{section_impacts}.
    \item[] Guidelines:
    \begin{itemize}
        \item The answer NA means that there is no societal impact of the work performed.
        \item If the authors answer NA or No, they should explain why their work has no societal impact or why the paper does not address societal impact.
        \item Examples of negative societal impacts include potential malicious or unintended uses (e.g., disinformation, generating fake profiles, surveillance), fairness considerations (e.g., deployment of technologies that could make decisions that unfairly impact specific groups), privacy considerations, and security considerations.
        \item The conference expects that many papers will be foundational research and not tied to particular applications, let alone deployments. However, if there is a direct path to any negative applications, the authors should point it out. For example, it is legitimate to point out that an improvement in the quality of generative models could be used to generate deepfakes for disinformation. On the other hand, it is not needed to point out that a generic algorithm for optimizing neural networks could enable people to train models that generate Deepfakes faster.
        \item The authors should consider possible harms that could arise when the technology is being used as intended and functioning correctly, harms that could arise when the technology is being used as intended but gives incorrect results, and harms following from (intentional or unintentional) misuse of the technology.
        \item If there are negative societal impacts, the authors could also discuss possible mitigation strategies (e.g., gated release of models, providing defenses in addition to attacks, mechanisms for monitoring misuse, mechanisms to monitor how a system learns from feedback over time, improving the efficiency and accessibility of ML).
    \end{itemize}

\item {\bf Safeguards}
    \item[] Question: Does the paper describe safeguards that have been put in place for responsible release of data or models that have a high risk for misuse (e.g., pretrained language models, image generators, or scraped datasets)?
    \item[] Answer: \answerNA{} 
    \item[] Justification: The paper poses no such risks.
    \item[] Guidelines:
    \begin{itemize}
        \item The answer NA means that the paper poses no such risks.
        \item Released models that have a high risk for misuse or dual-use should be released with necessary safeguards to allow for controlled use of the model, for example by requiring that users adhere to usage guidelines or restrictions to access the model or implementing safety filters. 
        \item Datasets that have been scraped from the Internet could pose safety risks. The authors should describe how they avoided releasing unsafe images.
        \item We recognize that providing effective safeguards is challenging, and many papers do not require this, but we encourage authors to take this into account and make a best faith effort.
    \end{itemize}

\item {\bf Licenses for existing assets}
    \item[] Question: Are the creators or original owners of assets (e.g., code, data, models), used in the paper, properly credited and are the license and terms of use explicitly mentioned and properly respected?
    \item[] Answer: \answerYes{} 
    \item[] Justification: The paper cites the original paper that produced the code package or dataset.
   \item[] Guidelines:
    \begin{itemize}
        \item The answer NA means that the paper does not use existing assets.
        \item The authors should cite the original paper that produced the code package or dataset.
        \item The authors should state which version of the asset is used and, if possible, include a URL.
        \item The name of the license (e.g., CC-BY 4.0) should be included for each asset.
        \item For scraped data from a particular source (e.g., website), the copyright and terms of service of that source should be provided.
        \item If assets are released, the license, copyright information, and terms of use in the package should be provided. For popular datasets, \url{paperswithcode.com/datasets} has curated licenses for some datasets. Their licensing guide can help determine the license of a dataset.
        \item For existing datasets that are re-packaged, both the original license and the license of the derived asset (if it has changed) should be provided.
        \item If this information is not available online, the authors are encouraged to reach out to the asset's creators.
    \end{itemize}
    
\item {\bf New Assets}
    \item[] Question: Are new assets introduced in the paper well documented and is the documentation provided alongside the assets?
    \item[] Answer: \answerYes{} 
    \item[] Justification: This paper relates the details of the code as part of the submission.
    \item[] Guidelines:
    \begin{itemize}
        \item The answer NA means that the paper does not release new assets.
        \item Researchers should communicate the details of the dataset/code/model as part of their submissions via structured templates. This includes details about training, license, limitations, etc. 
        \item The paper should discuss whether and how consent was obtained from people whose asset is used.
        \item At submission time, remember to anonymize your assets (if applicable). You can either create an anonymized URL or include an anonymized zip file.
    \end{itemize}

\item {\bf Crowdsourcing and Research with Human Subjects}
    \item[] Question: For crowdsourcing experiments and research with human subjects, does the paper include the full text of instructions given to participants and screenshots, if applicable, as well as details about compensation (if any)? 
    \item[] Answer: \answerNA{} 
    \item[] Justification: The paper does not involve crowdsourcing nor research with human subjects.
    \item[] Guidelines:
    \begin{itemize}
        \item The answer NA means that the paper does not involve crowdsourcing nor research with human subjects.
        \item Including this information in the supplemental material is fine, but if the main contribution of the paper involves human subjects, then as much detail as possible should be included in the main paper. 
        \item According to the NeurIPS Code of Ethics, workers involved in data collection, curation, or other labor should be paid at least the minimum wage in the country of the data collector. 
    \end{itemize}

\item {\bf Institutional Review Board (IRB) Approvals or Equivalent for Research with Human Subjects}
    \item[] Question: Does the paper describe potential risks incurred by study participants, whether such risks were disclosed to the subjects, and whether Institutional Review Board (IRB) approvals (or an equivalent approval/review based on the requirements of your country or institution) were obtained?
    \item[] Answer: \answerNA{} 
    \item[] Justification: The paper does not involve crowdsourcing nor research with human subjects.
    \item[] Guidelines:
    \begin{itemize}
        \item The answer NA means that the paper does not involve crowdsourcing nor research with human subjects.
        \item Depending on the country in which research is conducted, IRB approval (or equivalent) may be required for any human subjects research. If you obtained IRB approval, you should clearly state this in the paper. 
        \item We recognize that the procedures for this may vary significantly between institutions and locations, and we expect authors to adhere to the NeurIPS Code of Ethics and the guidelines for their institution. 
        \item For initial submissions, do not include any information that would break anonymity (if applicable), such as the institution conducting the review.
    \end{itemize}

\item {\bf Declaration of LLM usage}
    \item[] Question: Does the paper describe the usage of LLMs if it is an important, original, or non-standard component of the core methods in this research? Note that if the LLM is used only for writing, editing, or formatting purposes and does not impact the core methodology, scientific rigorousness, or originality of the research, declaration is not required.
    \item[] Answer:  \answerNA{} 
    \item[] Justification: The core method development in this research does not involve LLMs as any important, original, or non-standard components.
    \item[] Guidelines:
    \begin{itemize}
        \item The answer NA means that the core method development in this research does not involve LLMs as any important, original, or non-standard components.
        \item Please refer to our LLM policy (\url{https://neurips.cc/Conferences/2025/LLM}) for what should or should not be described.
    \end{itemize}
    
\end{enumerate}
\end{document}